\documentclass[10pt,a4paper,twoside]{aa}
		
\usepackage{graphicx}
\usepackage{latexsym}
\usepackage{amsmath}
\usepackage{amssymb}
\usepackage{amstext}
\usepackage{color}
\usepackage{psfrag}

\begin{document}

	\title{\textit{XMM-Newton} observations of the 
	\mbox{supernova remnant RX J1713.7--3946} and its central source}

	\titlerunning{\textit{XMM-Newton} observations of 
	SNR RX J1713.7--3946}

	\author{\textsc{G.~Cassam-Chena\"{i}}\inst{1} 
		\and \textsc{A.~Decourchelle}\inst{1}
		\and \textsc{J.~Ballet}\inst{1}
		\and \textsc{J.-L.~Sauvageot}\inst{1}
		\and \textsc{G.~Dubner}\inst{2}
		\and \textsc{E.~Giacani}\inst{2}}

	\authorrunning{\textsc{G.~Cassam-Chena\"{i}} et al.}

	\offprints{Gamil \textsc{Cassam-Chena\"{i}}, \\
	\email{gcc@discovery.saclay.cea.fr}}

	\institute{Service d'Astrophysique, CEA Saclay, 91191 
	Gif-sur-Yvette Cedex, France
		\and Instituto de Astronom\'{i}a y F\'{i}sica del Espacio, 
	CC 67, Suc. 28, 1428, Buenos Aires, Argentina}

\date{}

\abstract{We present new results coming from the observations of the supernova remnant (SNR) \object{RX J1713.7--3946} (also \object{G347.3--0.5})
performed in five distinct pointings with the EPIC instrument on board the satellite \textit{XMM-Newton}.
RX J1713.7--3946 is a shell-type SNR dominated by synchrotron radiation in the X-rays.
Its emission (emission measure and photon index) as well as the absorption along the line-of-sight has been characterized
over the entire SNR.
The X-ray mapping of the absorbing column density has revealed strong well constrained variations 
($0.4 \times 10^{22}$ cm$^{-2} \leq N_{\mathrm{H}} \leq 1.1 \times 10^{22}$ cm$^{-2}$)
and, particularly, a strong absorption in the southwest.
Moreover, there are several clues indicating that 
the shock front of RX J1713.7--3946 is impacting the clouds responsible for the absorption as revealed for instance by
the positive correlation between X-ray absorption and X-ray brightness along the western rims.
The CO and H\textsc{i} observations show that the inferred cumulative absorbing column densities 
are in excellent agreement with the X-ray findings in different places of the remnant
on condition that the SNR lies at a distance of $1.3 \pm 0.4$ kpc, probably in the Sagittarius galactic arm,
instead of the commonly accepted value of 6 kpc.
An excess in the CO emission is found in the southwest suggesting that the absorption is due to molecular clouds.
A search for OH masers in the southwestern region has been unsuccessful, possibly due to the low density of the clouds.
The X-ray mapping of the photon index has also revealed strong variations ($1.8 \leq \Gamma \leq 2.6$).
The spectrum is steep in the faint central regions and flat at the 
presumed shock locations, particularly in the southeast.
Nevertheless, the regions where the shock impacts molecular clouds have a steeper spectrum
than those where the shock propagates into a low density medium.
The search for the thermal emission in RX J1713.7--3946 has been still unsuccessfull leading
to a number density upper limit of $2 \times 10^{-2}$ cm$^{-3}$ in the ambient medium.
This low density corresponds to a reasonable kinetic energy of the explosion provided that the remnant is less than a few thousand years old.
A scenario based on a modified ambient medium due to the effect of a progenitor stellar wind is proposed and leads
to an estimate of RX J1713.7--3946's progenitor mass between 12 and $16 \: \mathrm{M}_{\odot}$.
The X-ray bright central point source \mbox{1WGA J1713.4--3949} detected at the center of SNR RX J1713.7--3946
shows spectral properties very similar to those of the Compact Central Objects found in SNRs 
and consistent in terms of absorption with that of the central diffuse X-ray emission arising from the SNR.
It is highly probable that the point source \mbox{1WGA J1713.4--3949}
is the compact relic of RX J1713.7--3946's supernova progenitor.
}

\maketitle

\section{Introduction}
Supernova remnants (SNRs) are well known
to be sources of radio synchrotron emission from GeV relativistic electrons.
During the last 10 years, X-ray synchrotron
emission from TeV relativistic electrons has been discovered in several 
shell-type SNRs among which are SN 1006 (Koyama et al. 1995), 
RX J1713.7--3946 (Koyama et al. 1997, Slane et al. 1999) 
and RX J0852.0--4622 (Slane et al. 2001).
This discovery bolstered the idea that collisionless 
shocks in SNRs are very powerful and efficient electron accelerators, 
possibly up to the "knee" of the cosmic-ray (CR) spectrum up to $10^3$ TeV.
It is reasonable to assume that protons are as well accelerated 
by these shocks.

RX J1713.7--3946 is a shell-type SNR located in the Galactic 
plane that was discovered with the \textit{ROSAT} all-sky survey in the 
constellation Scorpius (Pfeffermann \& Aschenbach 1996).
It has a slightly elliptical shape with a maximum extent of $70\arcmin$.
The observation of the northwestern shell of the SNR with the \textit{ASCA} 
satellite has shown the emission was exclusively non-thermal (Koyama et al. 1997). 

Further observations with \textit{ASCA} of most of 
the remnant did not reveal traces of the thermal emission, being likely 
overwhelmed by the bright X-ray synchrotron emission (Slane et al. 1999, 
hereafter SL99). 
An upper limit on the mean density of the ambient medium has been 
derived: $n_0 < 0.28 \; D_6^{-1/2}$ cm$^{-3}$, where $D_6$ is the 
distance in units of 6 kpc.
Recently, this thermal component may have been detected 
in the interior of the SNR over large fields-of-view using 
\textit{ROSAT}, \textit{ASCA} and \textit{RXTE} satellites 
(Pannuti et al. 2003).
However, some possible local variations of absorbing 
column density and photon index can compromise this interpretation.

An intriguing X-ray point-source (1WGA J1713.4--3949) with no optical 
and radio counterparts was also detected at the center of the remnant, being 
possibly a neutron star or an extra-galactic object (SL99).
Recently, a combined \textit{Chandra} and \textit{XMM-Newton} spectral 
analysis of this source has shown that its
X-ray properties
(two-component spectrum, luminosity, absence of pulsations) resemble
those of other compact central objects in SNRs and suggests 
that 1WGA J1713.4--3949 is the
neutron star associated to RX J1713.7--3946 (Lazendic et al. 2003).

Neither the distance nor the age of RX J1713.7--3946 is well established. 
From the X-ray measurements of the column density toward this source, 
Koyama et al. (1997) derived 
a distance of 1 kpc, while SL99 proposed a larger distance of 6 kpc 
based on its probable association with three dense and massive molecular 
clouds and the H\textsc{ii} region G347.6+0.2 located northwest to it.
However that may be, if the Sedov evolution is assumed with 6 kpc, 
an age of a few $10^4$ years results,
whereas an age of one order of magnitude below is compatible with 1 kpc.
Hence, this nearer distance could be in agreement 
with the hypothesis proposed by Wang et al. (1997),
based on historical records, that RX J1713.7--3946 is the 
remnant of the supernova (SN) that exploded in AD 393.
Following studies adopted the largest distance of 6 kpc, until
new \textit{XMM-Newton} observations (Cassam-Chena\"{i} et al. 2004b)
that re-opened the whole distance question together with
new high-resolution CO mm-wave observations (Fukui et al. 2003).
These new results suggest possible indications
of interaction between the SNR shock front and molecular gas located at 1 kpc
in the northwest and southwest sides of the SNR.

In the radio, the emission arises from faint filaments aligned with the 
X-ray shell of RX J1713.7--3946, except for the brightest filament located 
in the northwestern rim which is perpendicular to the shock and coincident 
with part of the H\textsc{ii} region G347.6+0.2 (SL99, Ellison et al. 2001, 
Lazendic et al. 2004). 

GeV $\gamma$-ray emission was detected by the \textit{EGRET} instrument
to the northeast of the SNR (Hartman et al. 1999). 
This emission was interpreted as the decay of neutral pions attributed 
to the interaction of CR nuclei (accelerated at the shock in RX J1713.7--3946)
with the massive and dense cloud assumed to be interacting with the remnant 
(Butt et al. 2001).
By arguing on the photon spectral index of the GeV source 
($\alpha = 2.3 \pm 0.2$ which is expected from the hadronic interactions of 
CR source) and on the non-detection in radio of the massive cloud
Butt et al. (2001) ruled out the possibility that electrons are 
responsible for the GeV luminosity.

At TeV $\gamma$-rays, the \textit{CANGAROO} imaging Cerenkov telescope 
detected emission in the northwest of the SNR, which was interpreted as 
Inverse Compton (IC) emission from accelerated electrons (Muraishi et al. 
2000). 
The low matter densities in the ambient medium is
unfavorable to the interpretation in terms of $\pi^0$ decay process.
A second \textit{CANGAROO} observation of RX J1713.7--3946 leads to a TeV 
$\gamma$-ray spectrum (between 400 GeV and 8 TeV) whose photon spectral index 
($\alpha = 2.8 \pm 0.2$) was shown to be consistent with the $\pi^0$ decay 
process and no other mechanism (Enomoto et al. 2002).
However, such a spectrum would exceed the \textit{EGRET} observed 
emission by a factor three (Reimer \& Pohl 2002, Butt et al. 2002).

Several self-consistent models have been constructed using the broad-band  
spectrum from radio to $\gamma$-ray wavelengths, essentially for the 
northwest region of RX J1713.7--3946. 
Using a non-linear diffusive shock acceleration model, Ellison et al. 
(2001) were able to reproduce the broadband spectrum with synchrotron 
emission from shock-accelerated electrons in X-rays and IC emission in 
$\gamma$-rays.
In this model, more than $25\%$ of the shock kinetic energy is going into 
relativistic ions. Other attempts to reproduce the broadband spectrum were 
undertaken (Uchiyama et al. 2003; Pannuti et al. 2003; Lazendic et al. 2004).

In this paper, we provide for the first time a detailed description of 
the X-ray emission of RX J1713.7--3946 based on \textit{XMM-Newton} observatory data.
Thanks to the large field-of-view provided by the EPIC imaging spectrometers, 
we can cover the bulk of the SNR extent in a few pointings. The high 
sensitivity of \textit{XMM-Newton} allows us to carry out a spectral 
analysis at medium-scale of the emission structures 
and then to produce for the 
first time a mapping of the spectral parameters of 
RX J1713.7--3946.

In Section \ref{data_processing}, we present our observations and the different data reduction methods that we have used.
In Section \ref{results}, we present the results obtained from the \textit{XMM-Newton} data analysis.
Section \ref{discussion} contains a discussion on 
the potential association between the central point-like source 1WGA J1713.4--3949 and SNR RX J1713.7--3946 (Sect. \ref{SNII}),
on the interaction between the remnant and clouds
as well as on the SNR distance (Sect. \ref{interaction_with_clouds}), on the SNR age and energetics
(Sect. \ref{energetics}) and finally on the overall picture that comes out from the results (Sect. \ref{evol_stage}).
In Section \ref{conclusion}, we summarize our results.

\section{Data processing}\label{data_processing}

\newsavebox{\blackbox}
\savebox{\blackbox}{\textcolor{black}{\rule{2mm}{2mm}}}
\begin{table*}[th]
\centering
\begin{tabular}{cccccccl}
Pointing & Symbol & OBS\_ID & Date & \multicolumn{3}{c}{Exposure time (ks)} & Comment\\ & & & & MOS1 & MOS2 & pn & \\\hline \hline
NE & $\Box$ & 0093670101 & 2001-09-05 & 1.6 & 2.0 & 0 & Very bad quality \\
NW & \usebox{\blackbox} & 0093670201 & 2001-09-05 & 4.0 & 5.4 & 1.0 & Bad quality \\
SW & {\large $\bullet$} & 0093670301 & 2001-09-08 & 15.3 & 15.5 & 10.0 & Good quality \\
SE & {\large $\circ$} & 0093670401 & 2002-03-14 & 12.2 & 12.2 & 6.8 &  Good quality \\
CE & $\triangle$ & 0093670501 & 2001-03-02 & 13.0 & 13.2 & 7.3 & Good quality \\ \hline \hline
\end{tabular}
\caption{\footnotesize Duration after flare rejection. The location of the different pointings is shown on Fig. \ref{X_ray_images}}.
\label{duree_obs}
\end{table*}

\subsection{Observation}
SNR RX J1713.7--3946 was observed with the three EPIC instruments (namely the MOS1, MOS2 and pn cameras) 
on board the \textit{XMM-Newton} satellite in the course of the AO-1 program.
The observations were performed in five distinct pointings, each of around 10 ks duration.
\mbox{Table \ref{duree_obs}} indicates the date of observation and the OBS\_ID for each pointing.
The five pointings were chosen based on the \textit{ROSAT} PSPC image (SL99) to allow a maximum coverage of the remnant
and the possibility to extract a local X-ray background.

The pointings correspond schematically to the center, northeast, northwest, southwest and southeast 
(hereafter CE, NE, NW, SW, SE, respectively) parts of RX J1713.7--3946.
During the observation, the MOS and pn cameras were operated in 
Full Frame and Extended Full Frame providing a temporal resolution of 2.6 s and 200 ms, respectively.
For all the pointings, the medium filter was used to avoid contamination
by visible sources contained in the field of view
and to limit the number of low energy photons (and hence pile-up) coming from the bright X-ray point source \mbox{1WGA J1713.4--3949}.

The potential neutron star \mbox{1WGA J1713.4--3949} is located
roughly in the middle of the SNR at the position 
$\alpha_{\mathrm{J2000}}=$17h13m28.4s, $\delta_{\mathrm{J2000}}=$-39$^{\circ}$49$\arcmin$54.5$\arcsec$ (Lazendic et al. 2003).
In the hope to detect thermal emission lines around 1 keV from \mbox{1WGA J1713.4--3949}, 
the RGS data were also processed in the central pointing.

\subsection{Flare rejection}
The Science Analysis System (SAS version 5.3) was used for data reduction.

First, we have generated calibrated events files with the SAS tasks \textit{emchain} and \textit{epchain}.
To create light curves, spectra and images, we have selected the single, double, triple and quadruple events 
(pattern$\leq$12) for the MOS cameras, and the singles and doubles (pattern$\leq$4) for the pn camera 
(with FLAG$=$0 for all cameras).
The data quality of the CE, SW, SE pointings is very good. 
However, it is not the case for the NE and NW pointings which are contamined by soft protons flares.
The method used to clean the data from the flares 
consists in applying a threshold on the light curves at high energy, \emph{i.e.} where the 
EPIC cameras are almost insensitive to X-ray photons. 
We took a threshold of 18 counts per 100 seconds time interval in the 10-12 keV band for the MOS cameras (pattern$\leq$12)
and a threshold of 22 counts per 100 seconds time intervals in the 12-14 keV band for the pn camera (pattern=0).
Table \ref{duree_obs} indicates the remaining exposure time after flares cleaning for each EPIC camera.

\subsection{Data reduction methods}
In this section, we describe the different methods that have been used to create the images, the hardness ratio maps,
the spectra and the mapping of the spectral parameters.

\subsubsection{Images}\label{images_method}
All the mosaiced images of the X-ray emission presented here are background subtracted, 
exposure and vignetting corrected (Sect. \ref{x_ray_morpho}). 
The background used for the subtraction is only instrumental and 
is estimated by using a reference observation (Read \& Ponman 2003)
which has been obtained with the same medium filter as in our observations.
No correction for the local astrophysical background is made because 
we have no information on it in the center of the remnant and further more
it strongly varies around the remnant which renders difficult any extrapolation towards the interior.
To correct for the exposure and the vignetting, we divide the
mosaiced background-subtracted count image by its mosaiced exposure map 
which includes the vignetting correction using the SAS task \textit{eexpmap}.
Each time, the mosaiced count image is adaptively smoothed with the SAS task \textit{asmooth}
and the resulting template is applied to smooth both the mosaiced background-subtracted count image
and its associated exposure map before the division.

We have not used the pn data for the intensity maps as it
introduces some problems of homogeneity 
in the correction for vignetting and exposure
due to the lack of pn data in the NE pointing (see Table \ref{duree_obs}).

\subsubsection{Hardness ratio maps}\label{hardness_method}
To create hardness ratio maps, we divide two background subtracted images made in different energy bands that
are previously adaptively smoothed. For the smoothing, the template of the image with the lowest statistics is always
applied to that with higher statistics. When we compute the ratio of the two maps, the statistical fluctuations can be amplified.
To limit this effect, the MOS and pn data are used, but
in return, the NE is missing in the hardness ratio maps (Sect. \ref{thermal_emission}).

\subsubsection{Spectra}\label{spectra_method}
All spectral analyses use the weight method described in Arnaud et al. (2001) to correct for the vignetting.
As for the background subtraction, we always use the double subtraction method described in Arnaud et al. (2002)
except for the mapping of the spectral parameters (see below, Sect. \ref{mapping_method}).
This method allows to take into account the difference, at low energy, between the local astrophysical background
of our observations and the one found in the reference observation.
This is done for local spectral analyses which require a precise knowledge of the
spectrum at low energy (Sect. \ref{thermal_emission}).
In our observations, there are two regions where the local astrophysical background can be
extracted with enough statistics: the SW and the SE.
Each time, we choose the nearest available local astrophysical background for each selected region.

The spectra are grouped so that each bin has a signal-to-noise ratio greater than 5$\sigma$.
To fit the spectra, we use XSPEC (version 11.2, Arnaud 1996).
The chosen on-axis response files are
\texttt{m$*$\_medv9q20t5r6\_all\_15.rsp} for MOS and 
\texttt{epn\_ef20\_sdY9\_medium.rsp} for pn.

\subsubsection{Mapping of the spectral parameters}\label{mapping_method}
\begin{figure*}[th]
\centering
\begin{tabular}{cc}
\includegraphics[width=9cm]{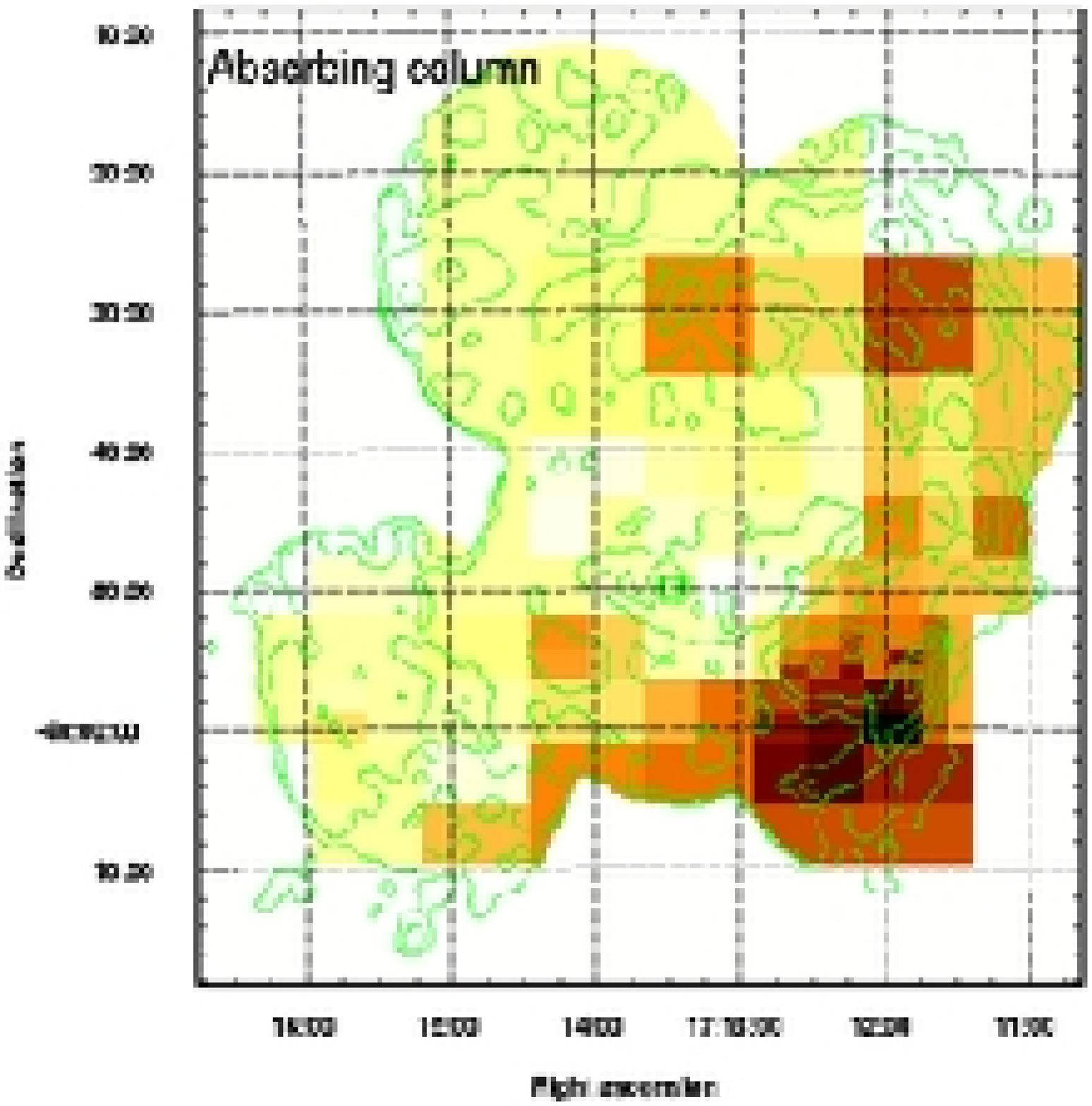} & \includegraphics[width=9cm]{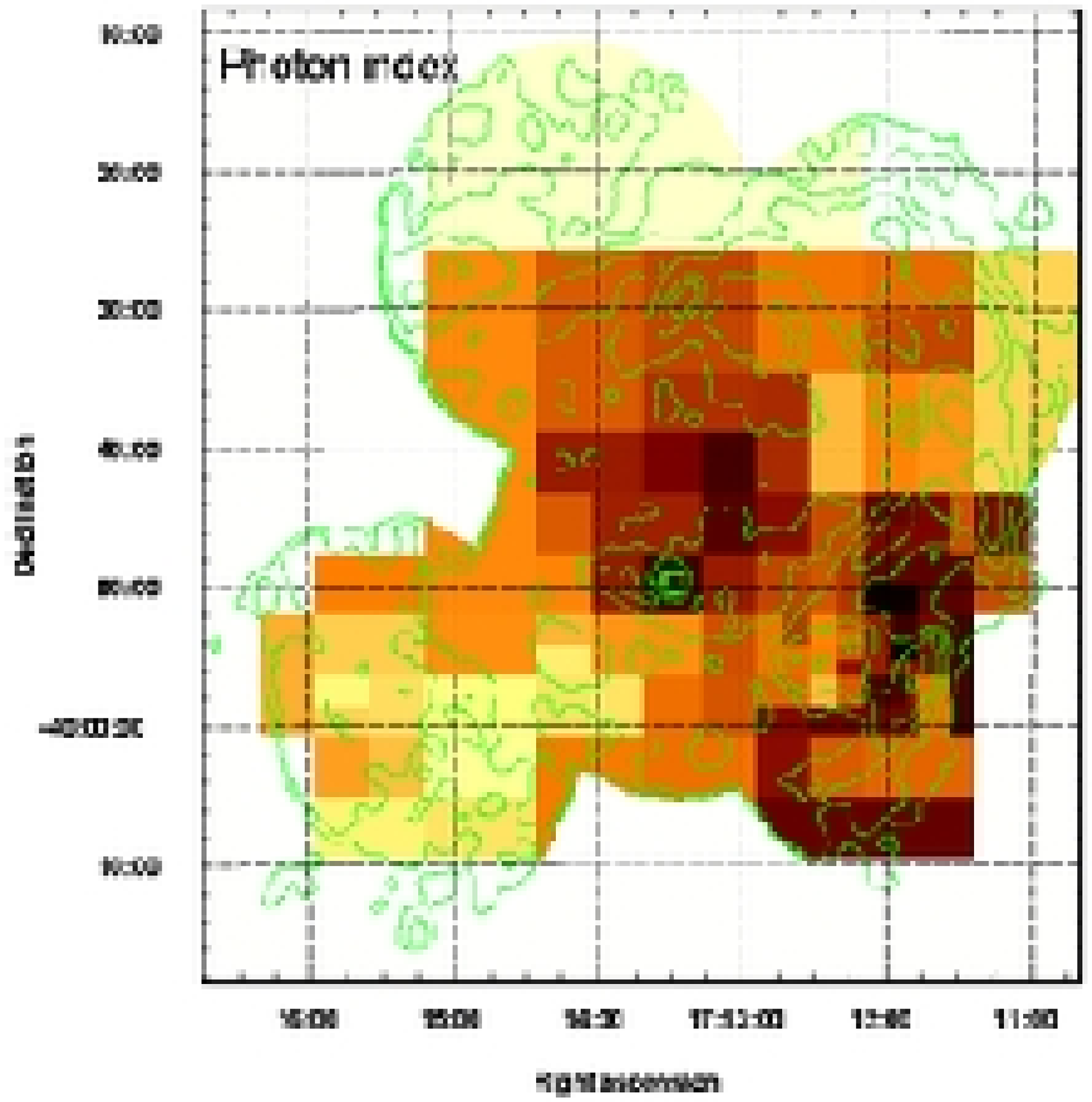} \\
\includegraphics[width=8cm]{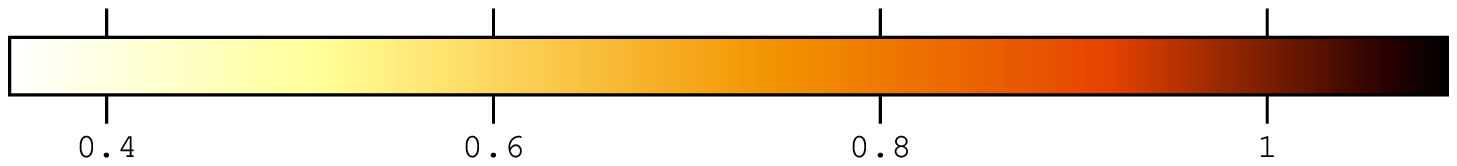} & \includegraphics[width=8cm]{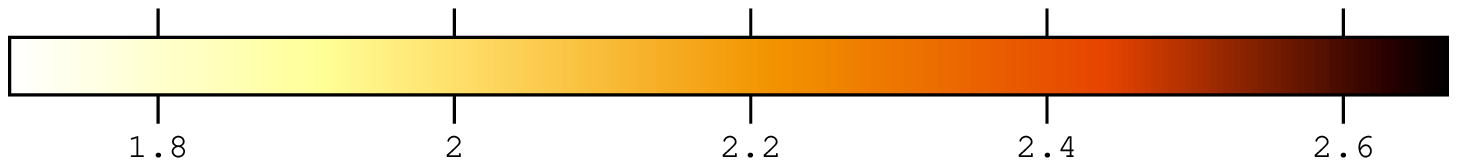} 
\end{tabular}
\caption{\textit{Left panel}: Absorbing column map $N_{\mathrm{H}}$ (in units of $10^{22}$ cm$^{-2}$) 
overlaid with the contour of the 2-10 keV image of SNR RX J1713.7--3946. 
The contour values start at $10^{-5}$ MOS counts/s/pixel and are square root 
spaced up to $1.9 \times 10^{-3}$ counts/s/pixel with a pixel size of $4\arcsec$.
The big boxes of our adaptive grid found in the NE and NW pointings are due to the low exposure time there.
\textit{Right panel}: Photon index map overlaid with the same X-ray contours. 
In both images, the scaling is linear.}
\label{maps}
\end{figure*}

The X-ray spectrum of SNR RX J1713.7--3946 
can be well fitted by
a simple absorbed power-law model the spectral parameters of which
are the absorbing column $N_{\mathrm{H}}$ and the photon index $\Gamma$.
Here, we describe how we map the spatial variations of these two parameters (Fig. \ref{maps}).

First, we build a spatial grid which takes into account 
the statistics so that we have approximately the 
same number of counts in each grid pixel.
The spatial grid is 
constructed from a mosaiced count image (MOS+pn) between 0.8-10 keV 
where all the point-like sources have been removed.
The mean number of total counts in a pixel is fixed to 9000 counts 
from this image which corresponds approximately to 
2500 and 4000 counts in most of the boxes of the MOS 
and pn cameras, respectively.

For each pixel of the grid, we extract the X-ray spectra (MOS, pn) 
and fit them with a simple absorbed power law.
A map of these parameters ($N_{\mathrm{H}}$, $\Gamma$) is constructed from the best-fit values in each grid pixel.

In this spectral analysis, we use a single background-subtracted method 
(as described for the images, Sect. \ref{images_method})
since it is not possible to have the true local astrophysical background 
in each grid pixel.
Because of that, the spectral bins only above 0.8 keV are selected, \emph{i.e.} where the 
reference background 
provided by Read \& Ponman (2003) is almost one order of magnitude below the spectrum 
of our observations.
We have checked that the absence of spectral bins below 0.8 keV did not 
change significantly the value of the spectral parameters in the X-ray bright regions.

There is no significant difference between the single and the double background-subtraction methods
for the brightest regions.
Nevertheless, in the X-ray faint regions, the effect of the double substraction is to
increase $N_{\mathrm{H}}$ and $\Gamma$ by a factor $\sim 10-15\%$ which is not
enough to alter our results and interpretation.

\begin{figure}[th]
\centering
\includegraphics[width=6cm,angle=-90]{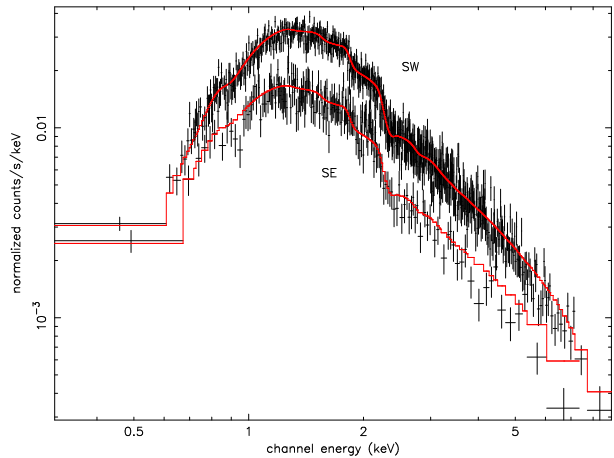}
\caption{EPIC pn spectra (double background subtraction) 
of the SW and SE regions of SNR RX J1713.7--3946. The fit with a power-law model gives
$N_{\mathrm{H}}=0.71 \pm 0.02 \times 10^{22}$ cm$^{-2}$ and $\Gamma = 2.33 \pm 0.03$ for the SW region and
$N_{\mathrm{H}}=0.57 \pm 0.03 \times 10^{22}$ cm$^{-2}$ and $\Gamma = 2.16 \pm 0.06$ for the SE region at a 90\% confidence level.}
\label{SW_SE_spectra_pn}
\end{figure}

\begin{figure*}[th]
\centering
\begin{tabular}{c}
\includegraphics[width=8cm]{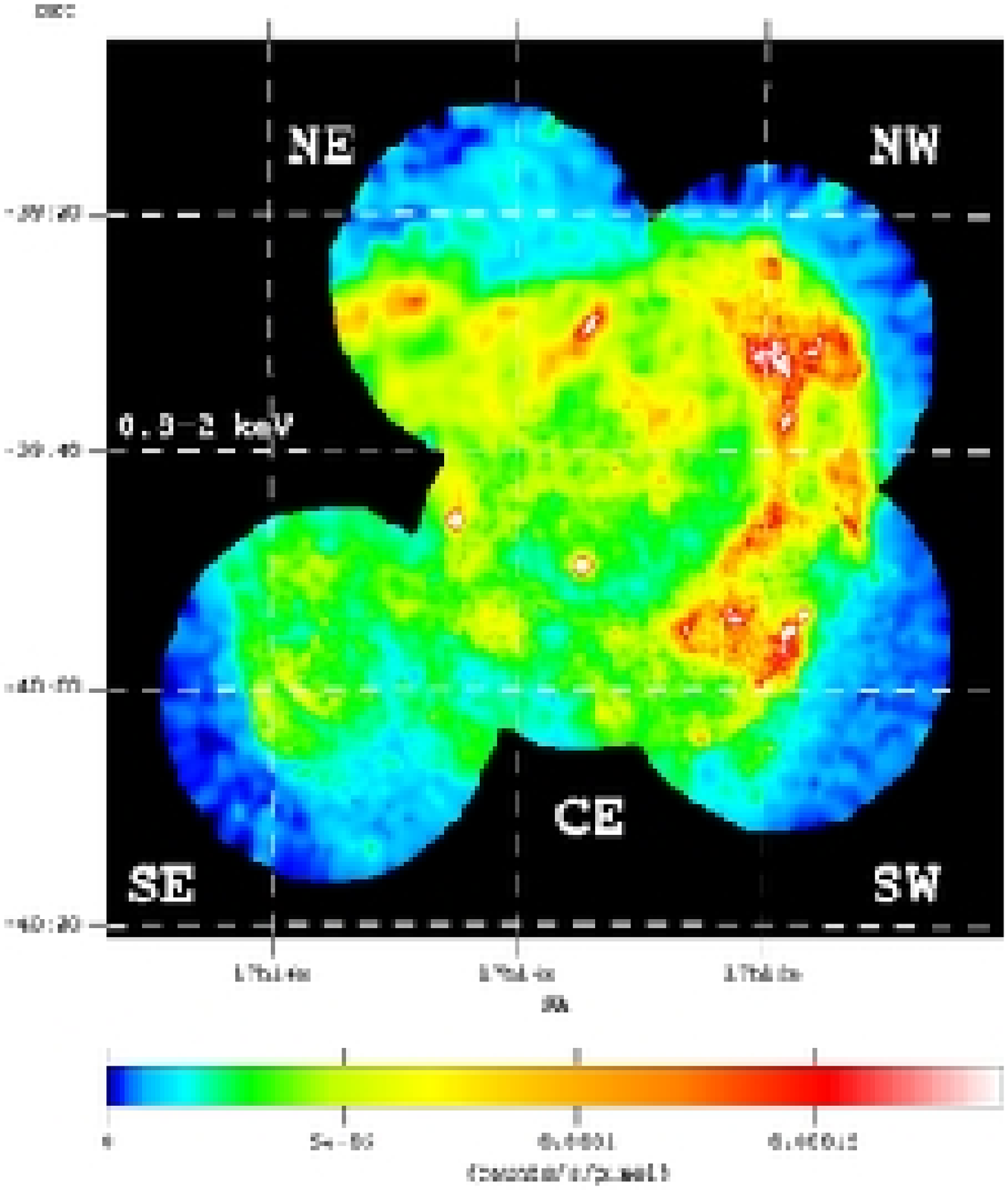}
\includegraphics[width=8cm]{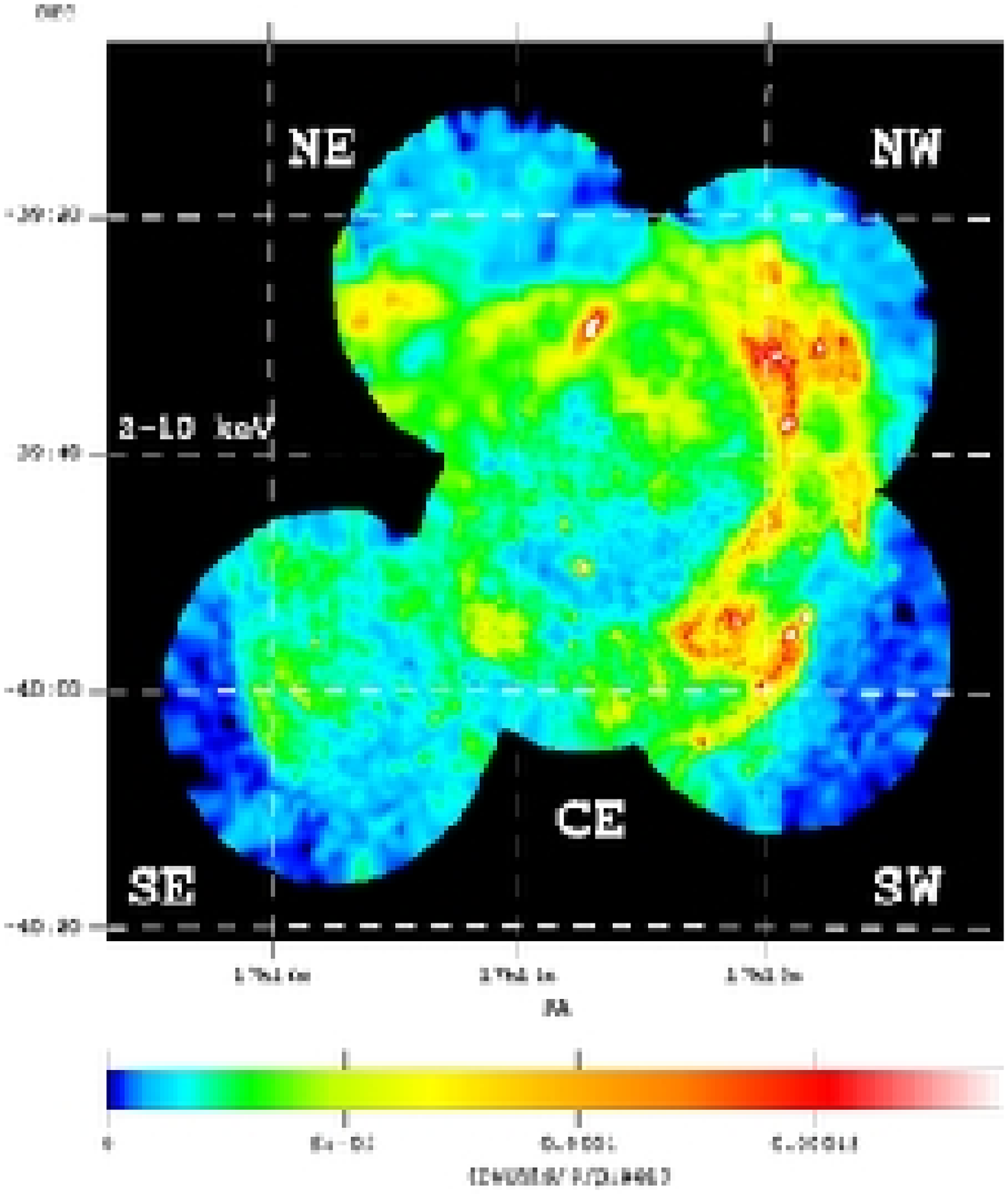}
\end{tabular}
\caption{\textit{Left panel}: MOS mosaiced image of SNR RX J1713.7--3946 
in the 0.8-2 keV energy band.
\textit{Right panel}: same as left panel but in the 2-10 keV energy band.
The neutron star \mbox{1WGA J1713.4--3949} is located in the middle 
of the CE pointing.
Both images were adaptively smoothed to a signal-to-noise ratio of 10. 
The scaling is square root with the higher cut fixed to 30\% 
of the SNR maximum X-ray brightness.
The color bar numbers are in units of counts/s/pixel with a pixel 
size of $4\arcsec$.
The bad resolution in the NE is due to a very short exposure time 
(see Table \ref{duree_obs}).
Moreover, it is possible that the smoothing has removed a few 
small-scale features in this NE pointing.}
\label{X_ray_images}
\end{figure*}

\begin{figure}[t]
\centering
\begin{tabular}{c}
\includegraphics[width=7cm]{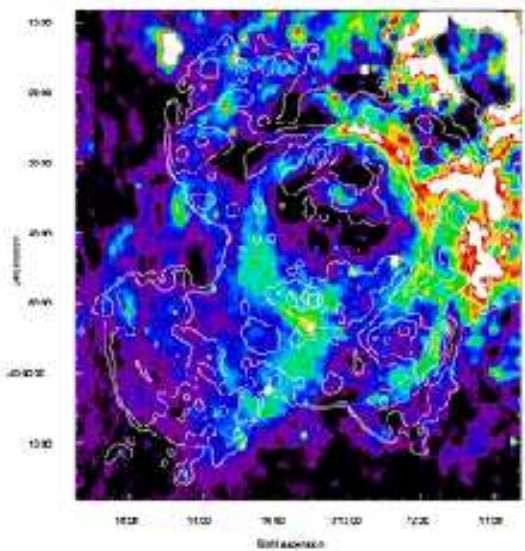}
\end{tabular}
\caption{1.4 GHz \textit{ATCA} radio image (courtesy of J.S. Lazendic)
overlaid with the 2-10 keV contour of SNR RX J1713.7--3946. 
The scaling is linear.
The contour values start at $10^{-5}$ MOS counts/s/pixel and are square root 
spaced up to $1.9 \times 10^{-3}$ counts/s/pixel with a pixel size of $4\arcsec$.}
\label{im_radio}
\end{figure}

\section{X-ray results}\label{results}

Unlike the X-ray spectra of typical shell-like SNRs which exhibit emission lines, 
the spectrum of RX J1713.7--3946 is found to be featureless 
as shown for example in Fig. \ref{SW_SE_spectra_pn}.
This suggests a non-thermal origin and 
has been interpreted as synchrotron X-ray emission 
from TeV shock accelerated electrons (SL99).
RX J1713.7--3946 is a second example, after SN 1006, of a shell-like 
SNR for which non-thermal emission dominates the X-ray flux, but
more extreme in the sense that no thermal emission has been unambiguously detected yet.
Unlike SN 1006, a symmetrical limb-brightened type Ia remnant 
located in a high-latitude environment,
RX J1713.7--3946 has a quite distorted morphology which may be
related to its complex surrounding environment in the Galactic plane and is likely to be the remnant
of a type II explosion as suggested by the presence of a central point source
detected at its center.

In section \ref{x_ray_morpho}, we describe in detail the unusual X-ray morphology of RX J1713.7--3946, 
and compare it to the radio continuum morphology.
In section \ref{syn_emis}, we trace and accurately diagnose the spectral variations of the 
synchrotron emission.
In section \ref{thermal_emission}, we search for the thermal emission and constraints on its properties.
Finally, in section \ref{Neutron_Star}, we intent to give better spectral constraints on the nature of
the point-like source \mbox{1WGA J1713.4--3949} to
provide arguments in favor or against a potential association with
SNR RX J1713.7--3946.

\subsection{X-ray morphology}\label{x_ray_morpho}
The good spatial resolution and sensitivity of \textit{XMM-Newton} combined to its large field-of-view
allow us for the first time to both have an overall view of RX J1713.7--3946 and resolve some X-ray emission features not seen before in the
\textit{ROSAT} and \textit{ASCA} images (SL99, Uchiyama et al. 2003).
Figure \ref{X_ray_images} shows the 5 pointings mosaiced image 
of RX J1713.7--3946 in the 0.8-2 keV (left panel)
and 2-10 keV (right panel) energy bands.
However, the 5 pointings are not enough for a complete coverage of the SNR.
The general shape of this distorted remnant 
resembles roughly an ellipse stretched out along the 
NW-SE axis over $\sim 70\arcmin$ corresponding to $\sim 20$ pc at 1 kpc or $\sim 120$ pc at 6 kpc.
An overall view of RX J1713.7--3946 obtained with \textit{ASCA} 
(Uchiyama et al. 2003) shows that the boundaries of
the SNR are irregular and suggests
that the remnant extends farther to the north.

Except for the point-like sources,
the comparison between the low energy and high energy images (see Fig. \ref{X_ray_images})
does not show significant changes in the X-ray morphology at first sight, 
at least for the brightest X-ray features.
From these images, the X-ray emission can be schematically classified in three distinct levels.
The lower intensity one is due to the diffuse X-ray emission of the local astrophysical background.
In both images, it can be found  (in blue-black) on the outside of the NE, NW, SW and SE pointings.
It stands out that the local astrophysical background intensity varies around the remnant.
This fact is expected since RX J1713.7--3946 lies just below the Galactic Ridge
in an environment where the properties of the interstellar gas rapidly change 
with Galactic latitude (which is the direction where the SNR is the more stretched out).
The medium intensity level corresponds to the diffuse and faint X-ray emission arising from the SNR which clearly 
dominates in the 0.8-2 keV energy band (in green and also in blue).
The higher intensity level is made of a more structured emission which appears in both 
images but more clearly at high energy in the 2-10 keV band (in yellow and red).
Its surface brightness is about twice higher than the diffuse emission associated with the SNR.

Both the NW and SW limbs are very bright in X-rays.
The X-ray emission in the western half is more structured than anywhere else in the SNR.
At large scale, a double-ring morphology stands out.
The internal ring is more or less circular whereas the external one looks broken into a few arcs.
At smaller scale, the \textit{Chandra} image of the NW region shows 
that the brightest X-ray features correspond to thin filaments of 
$20\arcsec$ apparent width or hot spots embedded in a 
diffuse plateau emission (Uchiyama et al. 2003, Lazendic et al. 2004).
This is confirmed by \textit{XMM-Newton} in the NW and also in the SW.

In spite of its bad quality, the NE observation shows that the inner circular thin shell seen in the west 
seems to continue there (see Fig. \ref{X_ray_images}, right panel).
Nevertheless, the diffuse X-ray emission from the NE pointing 
seems to exhibit a concave border between the 
remnant and the ambient medium (green emission in Fig. \ref{X_ray_images}, left panel).
Such a shape is not naturally expected from an 
explosion into a homogeneous low density interstellar medium.
That shape must be caused by some obstacles either interacting with
the SNR or obstructing the X-ray emission arising from it.

The X-ray emission from the SE is weak and diffuse but we distinguish even 
so a possible border between the SNR and the ambient medium.

Finally, the central part of the SNR shows diffuse X-ray emission 
rather weaker than in the NW and SW but similar to what is observed in the SE.
In addition, two point sources are observed.
The first one is \mbox{1 WGA J1714.4-3945} located at the position 
$\alpha_{\mathrm{J2000}}=$17h14m30s, $\delta_{\mathrm{J2000}}=$-39$^{\circ}$46$\arcmin$00$\arcsec$
which disappears above 2 keV. 
The second one located in the middle of the CE pointing is \mbox{1WGA J1713.4--3949}, 
a potential neutron star, which will be studied in detail in Section \ref{Neutron_Star}.

Figure \ref{im_radio} shows a high-resolution radio image 
obtained at 1.4 GHz with \textit{ATCA} (from Lazendic et al. 2004)
overlaid with the X-ray contours.
The radio morphology of RX J1713.7--3946 is
quite complex, with the brightest emission corresponding to the H\textsc{ii} region
G347.6+0.2, to the NW corner. A well defined ring of emission, about
15$\arcmin$ in size, is visible in the interior of the SNR. This radio feature
matches the inner X-ray bright ring, particularly towards the northern, northwestern and western
sides. Also, a very good correspondence is observed on the X-ray-bright outer rim to the SW.
The observed similar morphology in the radio and in the X-rays strongly suggests that 
the synchrotron emission detected at the two spectral domains must originate in the same 
electrons population accelerated at the shocks.

\begin{figure}[th]
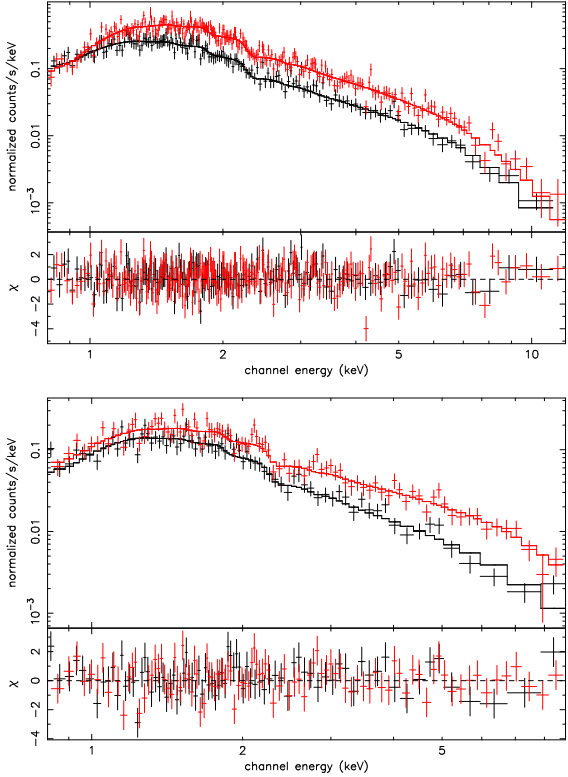

\centering
\begin{tabular}{c}
\includegraphics[width=5cm,angle=-90]{cassamchenai_9.ps}\\
\includegraphics[width=5cm,angle=-90]{cassamchenai_10.ps} 
\end{tabular} 
\caption{\textit{Top panel}: pn spectra of two SW regions of similar brightness ($\sim 2.8 \times 10^{-4}$ photons/cm$^2$/s/keV/arcmin$^2$) 
and same photon index ($\Gamma \sim 2.4$) but with different absorbing column densities.
The absorbing column is $1.09\pm{0.05} \times 10^{22}$ cm$^{-2}$ for the top spectrum 
(17h11m58.2s,-39$^{\circ}$59$\arcmin$44.5$\arcsec$)
and $0.77\pm{0.05} \times 10^{22}$ cm$^{-2}$ for the other one 
(17h11m52.5s,-39$^{\circ}$55$\arcmin$54.6.$\arcsec$).
\textit{Bottom panel}: pn spectra of two SW regions of similar brightness ($\sim 2 \times 10^{-4}$ photons/cm$^2$/s/keV/arcmin$^2$) 
and same absorbing column ($N_{\mathrm{H}} \sim 0.9 \times 10^{22}$ cm$^{-2}$) 
but with different photon index. 
$\Gamma = 2.07\pm{0.10}$ for the top spectrum 
(17h12m26.7s,-39$^{\circ}$57$\arcmin$34.4$\arcsec$)
and $\Gamma = 2.66\pm{0.12}$ for the other one
(17h11m52.6s,-39$^{\circ}$54$\arcmin$49.0$\arcsec$)} 
\label{sp_boxes}
\end{figure}

\begin{figure}[b]
\centering
\begin{tabular}{c}
\includegraphics[width=8cm]{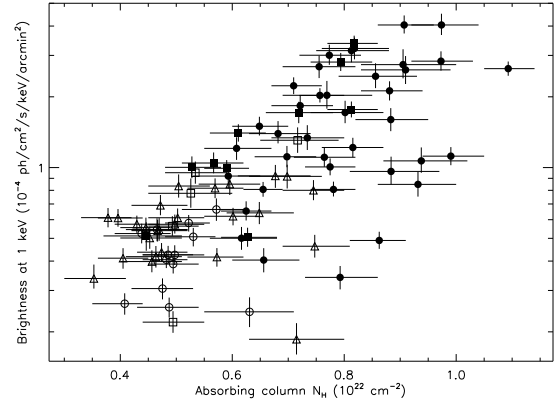}
\end{tabular}
\caption{Correlation plot between the absorbing column 
$N_{\mathrm{H}}$ (in units of $10^{22}$ cm$^{-2}$) and the normalized
brightness at 1 keV. 
Each point corresponds to the value found for a particular pixel of the mapping grid 
(Sect. \ref{mapping_method}) and
the error bars are given at a $90\%$ confidence level.
The symbols are defined as follow
(cf. Table \ref{duree_obs}):
$\Box$ for NE, \usebox{\blackbox} for NW, {\large $\bullet$} for SW, 
{\large $\circ$} for SE, $\triangle$ for CE.
The regions which are the more absorbed are the brightest.
They are mainly located in the SW.
The regions which have small $N_{\mathrm{H}}$ are the less bright and are found in the CE and SE.}
\label{correl_nh_br}
\end{figure}

\subsection{Spatial and spectral characterization of the synchrotron emission}\label{syn_emis}
In this section, we present the maps of absorbing column density and photon index
(method described in Sect. \ref{mapping_method}).

Figure \ref{maps} (left panel) shows the mapping of absorbing column 
density in the line-of-sight at medium scale.
The mean relative error on the absorbing column in each pixel 
grid is $\sim 9\%$ with a maximum value of $16\%$.
The variations of absorbing column density are strong 
with $N_{\mathrm{H}}$ varying from $\sim 0.4 \times 10^{22}$ cm$^{-2}$ to 
$\sim 1.1 \times 10^{22}$ cm$^{-2}$.
In the SE and CE pointings, $N_{\mathrm{H}}$ is low with a value $\sim 0.4-0.5 \times 10^{22}$ cm$^{-2}$ 
whereas it is larger in the NW ($N_{\mathrm{H}} \sim 0.6-0.7 \times 10^{22}$ cm$^{-2}$) 
and even more in the SW ($N_{\mathrm{H}} \sim 0.8-1.1 \times 10^{22}$ cm$^{-2}$).
Figure \ref{sp_boxes} (top panel) illustrates the variations of 
absorbing column from two spectra extracted in the SW region.

The comparison between the mapping of absorbing column 
density and the X-ray brightness contours
(Fig. \ref{maps}, left panel) suggests that there is
a positive correlation them.
Figure \ref{correl_nh_br} shows the X-ray brightness at 1 keV
versus $N_{\mathrm{H}}$ in each grid pixel.
We see that the regions of large (small) $N_{\mathrm{H}}$ 
correspond to those of high (weak) X-ray brightness
confirming the significance of the observed correlation. 
The fact that high absorbing column coincides with brighter
X-ray emission, as observed to the SW (and NW), is unexpected.
Indeed, a dense cloud located in front of an X-ray source 
and not related to it must diminish the photon source flux in our direction (at low energy in particular).
That we observe a positive
correlation between the absorbing column density and the X-ray brightness
means that the increased ambient density somehow amplifies the X-ray brightness.
This strongly suggests that the remnant is interacting in the brightest regions
with part of the line-of-sight absorbing material.

\begin{figure}[t]
\centering
\begin{tabular}{c}
\includegraphics[width=7cm]{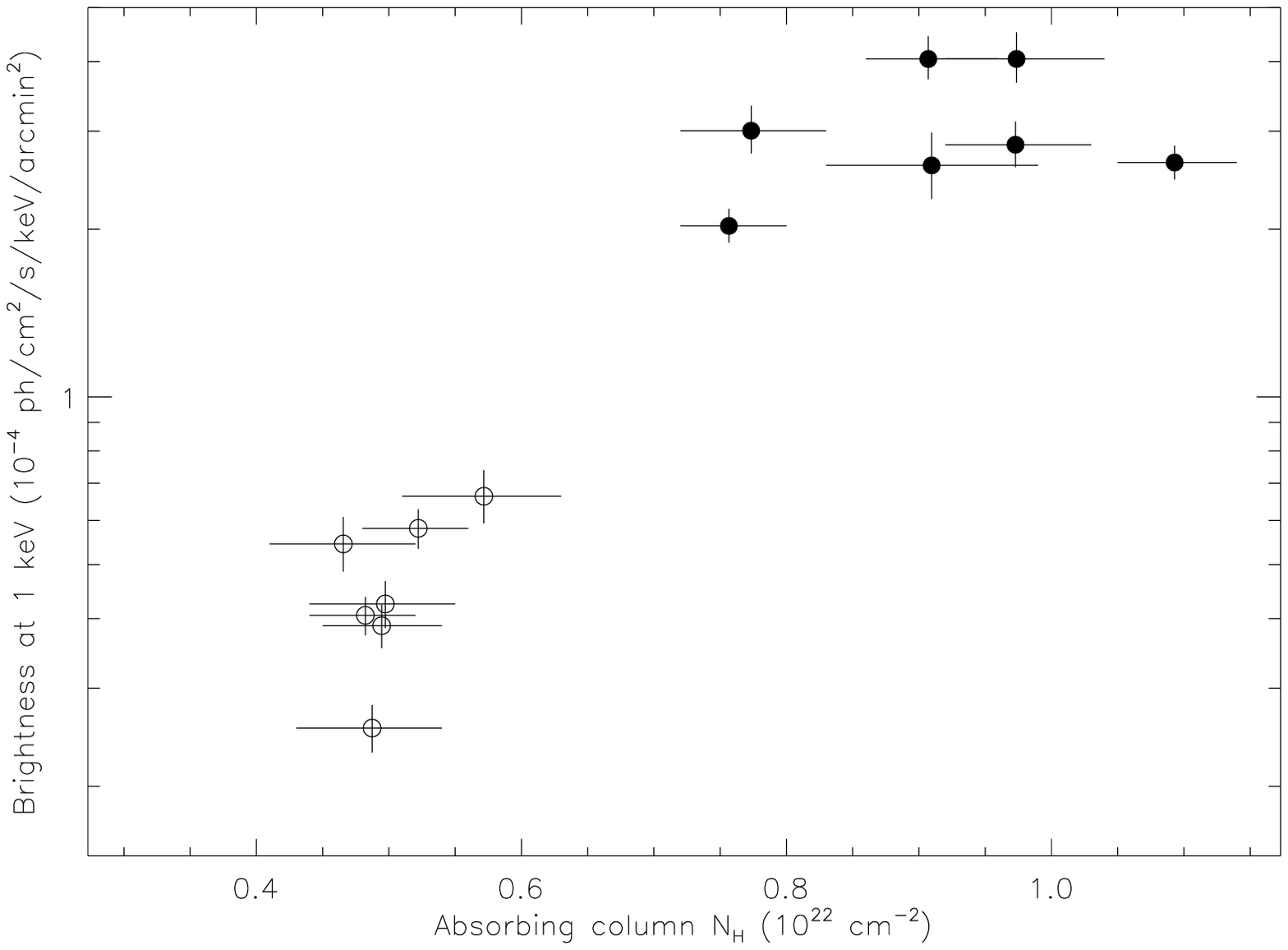} \\
\includegraphics[width=7cm]{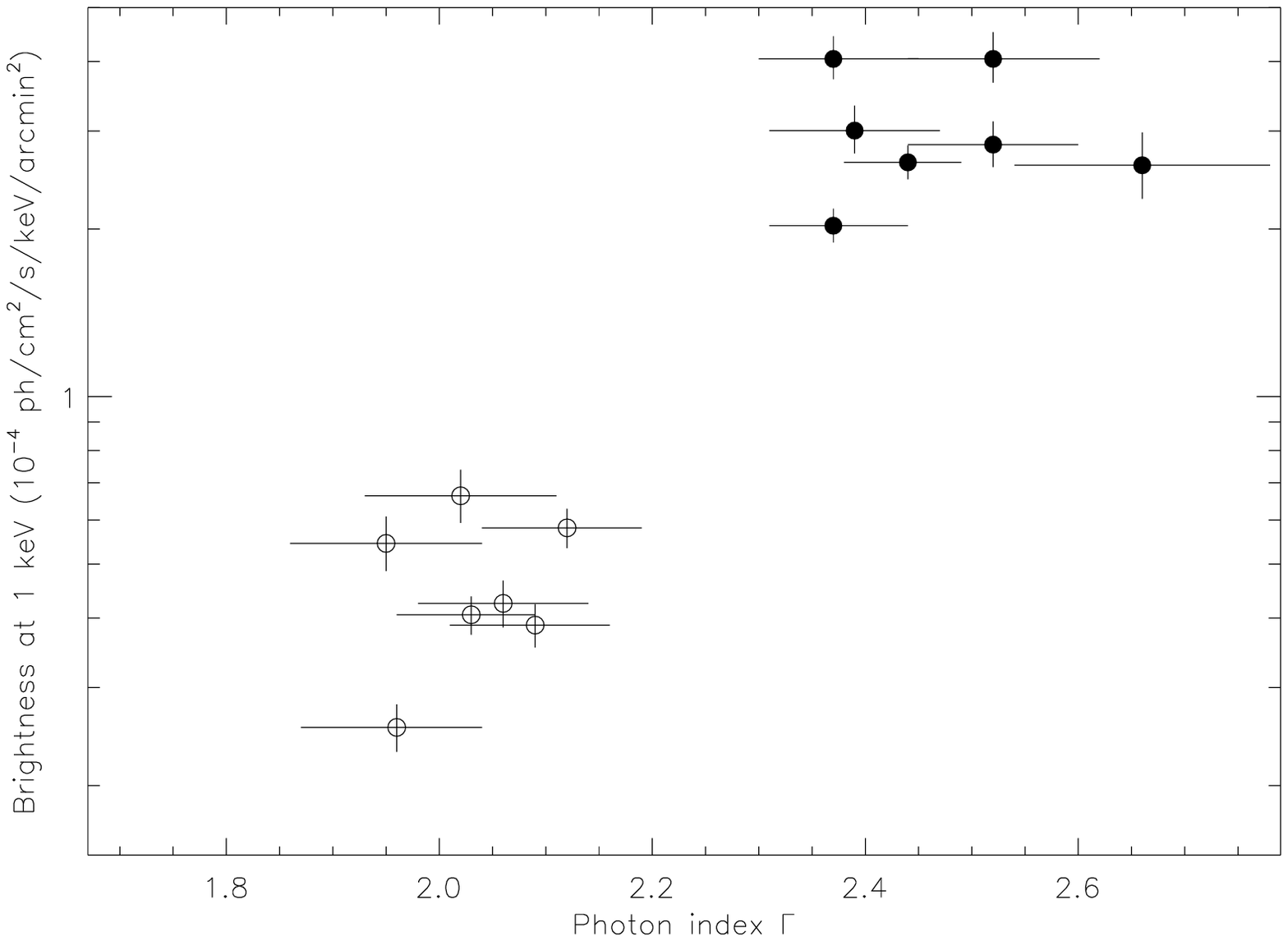}
\end{tabular}
\caption{\textit{Top panel}: Correlation plots at the shock location 
in the SW ({\large $\bullet$}) and SE ({\large $\circ$}) regions between 
the absorbing column 
$N_{\mathrm{H}}$ (in units of $10^{22}$ cm$^{-2}$) and the normalized
brightness at 1 keV (in units of $10^{-4}$ photons/cm$^2$/s/keV/arcmin$^2$).
Each point corresponds to the value found for a 
particular pixel of the mapping grid 
(Sect. \ref{mapping_method}) and
the error bars are given at a $90\%$ confidence level.
The points of the SW region are selected so that they are 
above $2.0 \times 10^{-4}$ photons/cm$^2$/s/keV/arcmin$^2$.
\textit{Bottom panel}: 
Correlation plots between the photon index $\Gamma$ and 
the normalized brightness at 1 keV for the same set of points as above. 
The error bars are given at a $90\%$ confidence level. }
\label{correl_nh_bri_select}
\end{figure}

Figure \ref{maps} (right panel) shows the mapping of the photon spectral index.
The mean relative error on the photon index value in each grid pixel is 
$\sim 4\%$ with a maximum value of $7\%$.
The variations of photon index are important 
seeing that $\Gamma$ varies from $\sim 1.8$ to $\sim 2.6$.
The spectrum is steep in the faint central region whereas 
it is a bit flatter in the NW and even more in the SE.
The  fact that the spectrum gets steeper as we move inwards 
away from the shock front is expected in SNRs. 
This effect is generally explained by the
adiabatic expansion of the remnant and the synchrotron 
radiative losses (Reynolds 1998) as is also observed in SN 1006.
In the SW, the spectrum is generally steep, but some regions exhibit strong variations of spectral index
as is shown in Fig. \ref{maps} (right panel).
For illustration,  Fig. \ref{sp_boxes} (bottom panel) shows the spectrum of 
two regions in the SW, for which the index is very different.

The synchrotron X-ray morphology, though quite complex, 
can be related to the absorption along the line-of-sight (or equivalently brightness).
On the one hand, there is a weakly absorbed SE region which seems to extend far away from
the center of the SNR (taken as the point-like source \mbox{1WGA J1713.4--3949}),
and on the other hand, a strongly absorbed western region which is closer
and where RX J1713.7--3946 is suspected to interact with the material that makes the absorption.
Such an asymmetry in the morphology and the absorption
must be the result of different ambient densities around the remnant.
To investigate how particle acceleration is modified in different environment,
we have selected two sets of points located at the presumed shock locations:
the one in the SW external rim coincident 
with the radio emission (see Fig. \ref{im_radio}) and in the SE.
These points are shown in Fig. \ref{correl_nh_bri_select} in the
absorbing column density-X-ray brightness plane (top panel) and
in the photon index-X-ray brightness plane (bottom panel).
It is obvious that two groups come out.
The regions where the shock propagates in the SW have a steep spectrum
($\Gamma \sim 2.3-2.7$) and those where the shock propagates
in the SE have a flat spectrum ($\Gamma \sim 1.9-2.2$)\footnote{The effect of the double 
subtraction method (cf. Sect. \ref{mapping_method})
is to tighten the spectral index values between the SW and the SE.
However the two distinct groups remain.}.
This point is discussed further in Sect.\ref{acceleration}.

\begin{figure*}[th]
\centering
\begin{tabular}{cc}
\includegraphics[width=8.5cm]{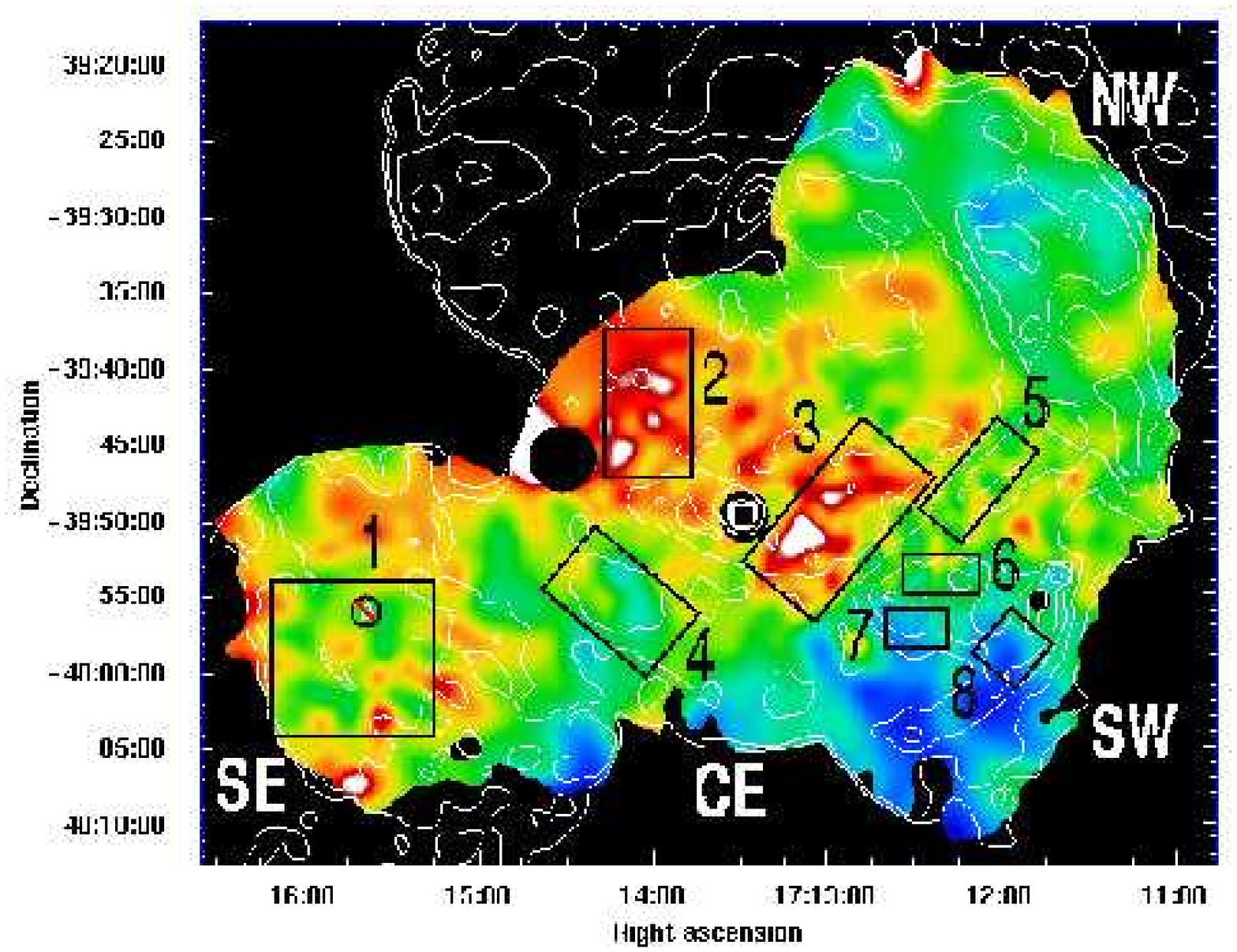} &
\includegraphics[width=8.5cm]{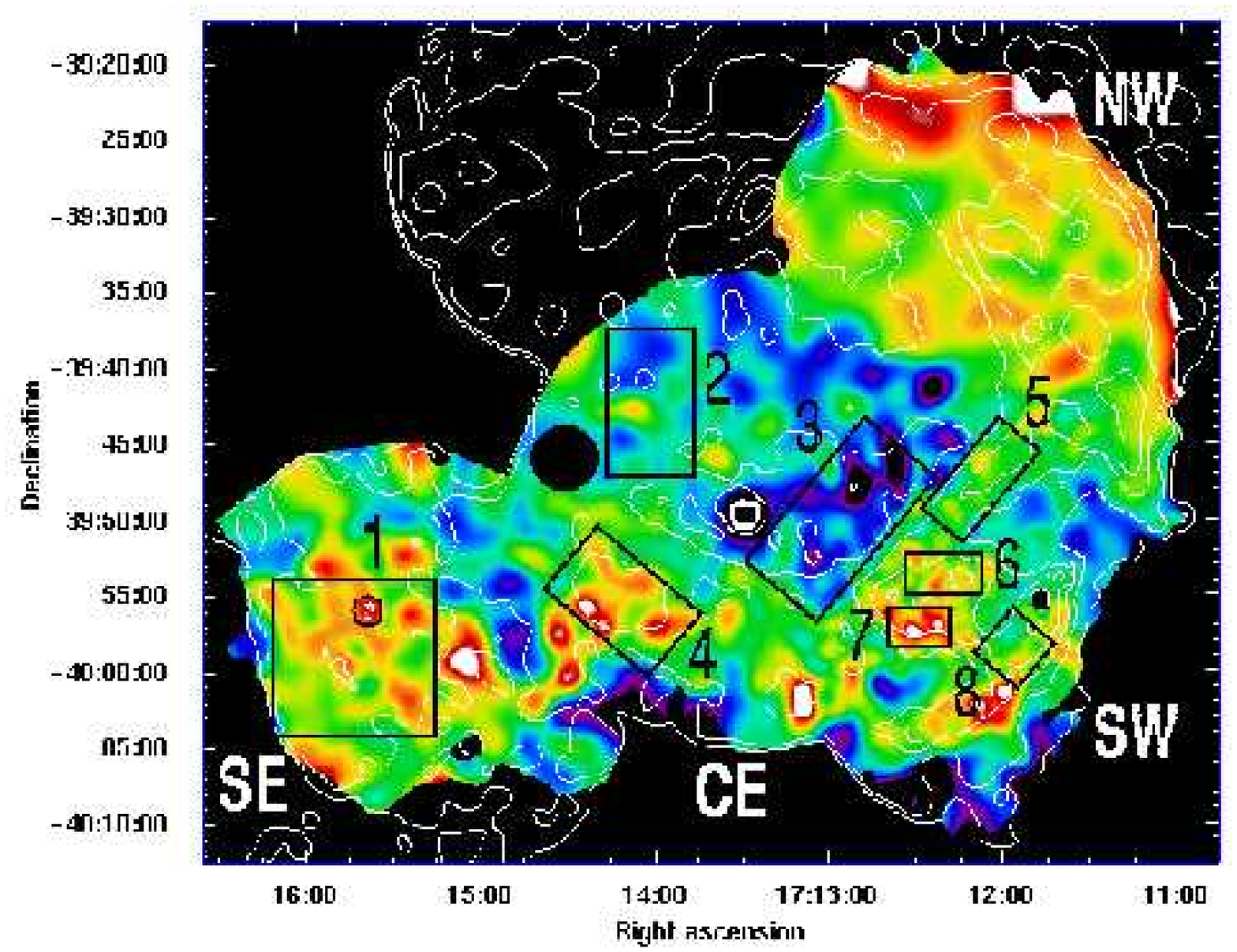} \\
\includegraphics[width=8cm,height=1cm]{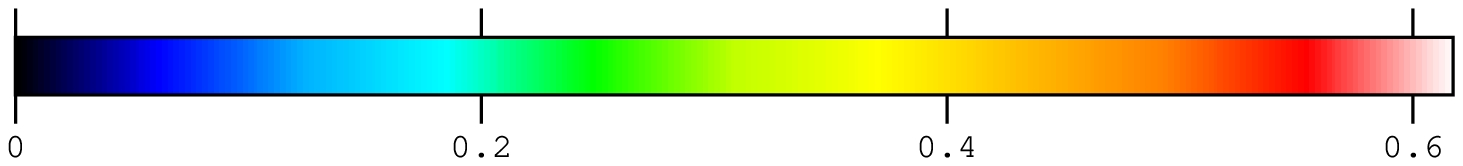} & 
\includegraphics[width=8cm,height=1cm]{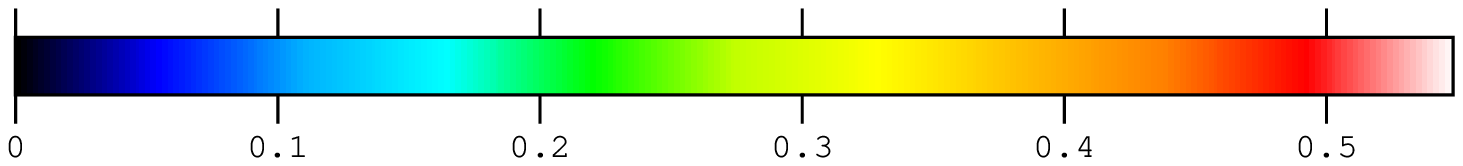}
\end{tabular}
\caption{\textit{Left panel}: Spatial distribution of the hardness ratio [0.7-0.9 keV]/[1.0-1.3 keV] 
overlaid with the contour of the 2-10 keV image of SNR RX J1713.7--3946.
\textit{Right panel}: Spatial distribution of the hardness ratio [5-8 keV]/[1.6-2.0 keV] overlaid with the same X-ray contours. 
In both images, the point-like sources have been removed. The scaling is linear.
Regions of extraction used for the search of the thermal emission are shown and labeled from 1 to 8.}
\label{hardness_ratio}
\end{figure*}

\begin{table*}[t]
\centering
\begin{tabular}{cllll}\hline \hline
Region & \multicolumn{1}{c}{$\chi^2$ (dof)} & \multicolumn{1}{c}{$N_{\mathrm{H}}$ ($10^{22}$ cm$^{-2}$)} & \multicolumn{1}{c}{$\Gamma$} & \multicolumn{1}{c}{$Norm$} \\ \hline
1 & 952 (909) & 0.66 (0.64-0.69) & 2.34 (2.29-2.39) & 6.06e-5 (5.75-6.41) \\ 
2 & 530 (468) & 0.54 (0.51-0.56) & 2.68 (2.61-2.75) & 6.94e-5 (6.51-7.39) \\
3 & 533 (495) & 0.47 (0.45-0.49) & 2.58 (2.51-2.65) & 3.59e-5 (3.36-3.86) \\
4 & 540 (518) & 0.77 (0.74-0.82) & 2.23 (2.17-2.29) & 8.33e-5 (7.73-8.93) \\
5 & 656 (632) & 0.68 (0.66-0.70) & 2.46 (2.42-2.51) & 1.65-e4 (1.58-1.73) \\
6 & 738 (642) & 0.77 (0.75-0.80) & 2.30 (2.26-2.35) & 2.26e-4 (2.15-2.38) \\
7 & 394 (356) & 0.89 (0.85-0.94) & 2.15 (2.09-2.22) & 1.54e-4 (1.42-1.67) \\
8 & 822 (761) & 1.05 (1.02-1.08) & 2.52 (2.48-2.56) & 3.04e-4 (2.90-3.20) \\ \hline
\end{tabular}
\caption{Best fit parameters for the power-law model using the MOS and pn data all at once.
The errors are in the range $\Delta \chi^2 < 2.7$ (90\% confidence level) on one parameter.
The regions shown in Fig. \ref{hardness_ratio}.
$Norm$ is given in units of photons/cm$^2$/s/keV/arcmin$^2$ at 1 keV.}
\label{mo_wapo_mp}
\end{table*}

\begin{table*}[t]
\centering
\begin{tabular}{cllllll}\hline \hline
Region & \multicolumn{1}{c}{$\chi^2$ (dof)} & \multicolumn{1}{c}{$N_{\mathrm{H}}$ ($10^{22}$ cm$^{-2}$)} & \multicolumn{1}{c}{$\Gamma$} & \multicolumn{1}{c}{$Norm$} & \multicolumn{1}{c}{$kT_{\mathrm{e}}$ (keV)} & \multicolumn{1}{c}{$\int n_{\mathrm{e}} n_{\mathrm{H}} \: dl$ (cm$^{-5}$)} \\ \hline 
1 & 933 (907) & 0.77 (0.74-0.80) & 2.45 (2.39-2.52) & 7.00e-5 (6.51-7.31) & 0.48 (0.36-0.60) & 8.75e15 (5.76-12.30) \\
2 & 504 (466) & 0.70 (0.66-0.75) & 2.83 (2.75-2.94) & 8.43e-5 (8.07-9.49) & 0.61 (0.58-0.66) & 2.55e16 (1.77-3.48) \\
3 & 499 (493) & 0.57 (0.53-0.60) & 2.59 (2.51-2.70) & 3.73e-5 (3.49-4.22) & 0.76 (0.70-0.83) & 1.72e16 (1.25-2.33)  \\
4 & 532 (516) & 0.92 (0.87-0.96) & 2.34 (2.27-2.42) & 9.67e-5 (8.59-9.97) & 0.79 (0.62-1.01) & 1.27e16 (0.53-2.00) \\
5 & 646 (630) & 0.75 (0.73-0.77) & 2.52 (2.47-2.56) & 1.80e-4 (1.74-1.99) & 0.54 (0.42-0.66) & 1.45e16 (0.76-2.64) \\
6 & 717 (640) & 0.89 (0.85-0.92) & 2.40 (2.35-2.45) & 2.59e-4 (2.45-2.77) & 0.61 (0.47-0.69) & 2.45e16 (1.56-3.48) \\
7 & 383 (354) & 0.95 (0.89-1.02) & 2.20 (2.12-2.27) & 1.64e-4 (1.50-1.80) & 0.28 ($\leq 0.45$) & 1.08e16 (0.55-1.65) \\
8 & 807 (759) & 1.13 (1.09-1.18) & 2.58 (2.53-2.63) & 3.32e-4 (3.12-3.54) & 0.61 (0.43-0.73) & 1.29e16 (0.66-1.82) \\ \hline 
\end{tabular}
\caption{Best fit parameters for the power-law plus equilibrium model using the MOS and pn data all at once.
The errors are in the range $\Delta \chi^2 < 2.7$ (90\% confidence level) on one parameter.
The regions are shown in Fig. \ref{hardness_ratio}.
$Norm$ is given in units of photons/cm$^2$/s/keV/arcmin$^2$ at 1 keV.}
\label{mo_wapoequil_mp}
\end{table*}

\begin{figure}[t]
\centering
\includegraphics[width=9cm]{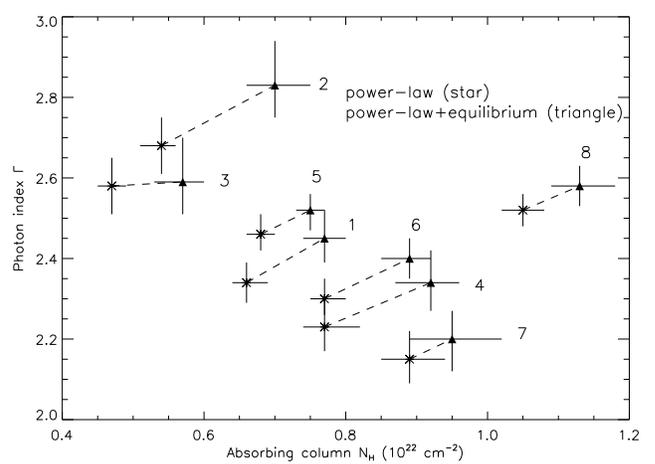}
\caption{For each of the 8 regions considered (see Fig. \ref{hardness_ratio}),
changes of the parameters ($N_{\mathrm{H}}$, $\Gamma$) with the addition of a thermal component.
The star correspond to the power-law model (cf. Table \ref{mo_wapo_mp}).
The black triangle correspond to the power-law plus thermal equilibrium model (cf. Table \ref{mo_wapoequil_mp}).
The error bars are given at a $90\%$ confidence level.}
\label{param_evol}
\end{figure}

\subsection{Thermal emission}\label{thermal_emission}

Determining the thermal properties of the hot gas is crucial 
to constrain the evolutionary state of the remnant.
That the thermal emission is not clearly detected in SNR RX J1713.7--3946
sets however strong limits on the temperature of the hot gas and, above all,
on its mean density. 
On the one hand, the temperature determination is important 
because it gives an idea on the shock velocity
(under a few assumptions on the heating of the electrons behind the shock).
On the other hand, the estimation of the mean ambient medium density is crucial to
provide an estimate of the swept-up mass (assuming that the distance to
the SNR is known) which then constrains the phase of evolution of the remnant.

To characterize the synchrotron emission (Sect. \ref{syn_emis}), the spectra of 
SNR RX J1713.7--3946 were modelled by an absorbed phenomenological power-law
which provides, at the first order, an excellent representation of the data.
To search for the thermal emission, we add a second component, the thermal model,
which will not necessarily improve the spectral fit but will be essential 
to constrain the properties of the thermal gas.
For this task, it is important to look at spectrally homogeneous areas
since the spectrum of a region with strong variations of $N_{\mathrm{H}}$ 
and $\Gamma$ modelled with an uniformly absorbed single power-law could simulate artificially the need 
for a thermal additional component instead of a second non-thermal component.
Eight particular regions (point-like sources removed) with no or few spectral variations were chosen based on the 
low-energy and high-energy hardness ratio maps, and such that they
have enough statistics. They are shown in Fig. \ref{hardness_ratio}.
Although the hardness ratio maps do not give the value of $\Gamma$ or $N_{\mathrm{H}}$,
they permit to keep a more precise spatial information than 
in the photon index or absorbing column map which 
intrinsically depend on the shape and size of the pixels of the grid.

Like in previous studies where an attempt to investigate 
the thermal emission properties was done (SL99, Pannuti et al. 2003),
we will first assume the plasma to be in an equilibrium ionization state.
This hypothesis will be discussed further.
In the following study, the elemental abundances are fixed to solar.

First, we investigate how our previous results (Sect. \ref{syn_emis}) might
be modified by the addition of a thermal component in the model.
Table \ref{mo_wapo_mp} gives the best-fit parameters obtained for an absorbed power-law model
in each of the eight regions selected for this analysis, while Table \ref{mo_wapoequil_mp}
gives the results obtained with the addition of a thermal equilibrium ionization component.
It appears that the introduction of the thermal model 
improves very slightly the fit
($\Delta \chi^2 \sim 10-30$) and does not change 
qualitatively the results found for
the non-thermal emission as is shown in Fig. \ref{param_evol}.
The derived parameters ($N_{\mathrm{H}}$, $\Gamma$) are similarly changed for all the eight regions:
they always get slightly larger as we add the thermal component to the model.

The comparison between the fluxes of the thermal emission obtained in different regions of the SNR gives
an idea on the spatial distribution of the shocked medium density.
Table \ref{mo_wapoequil_mp} gives the value of $\int n_{\mathrm{e}} n_{\mathrm{H}} \: dl$
and the temperature found for the thermal emission (for the 8 regions).
The range of temperatures $kT_{\mathrm{e}}$ is always below $1$ keV.
The values of the square density integrated along the line-of-sight are rather homogeneous ($\sim 1 \times 10^{16}$ cm$^{-5}$).
We note however that this quantity seems to increase from the
SNR shock front (regions 1 and 8) to the interior (regions 2 and 6) when we look at a temperature around 0.5-0.6 keV.

\begin{table}[t]
\centering
\begin{tabular}{ccc} \hline \hline
Region & \multicolumn{1}{c}{$\theta_{\mathrm{r}}$ ($\arcmin$)} & \multicolumn{1}{c}{$\hat{n}_{\mathrm{e}}$ ($10^{-2} D_{1}^{-1/2}$ cm$^{-3}$)}  \\ \hline
1 & 28 & 1.95 (1.58-2.31) \\
2 & 9 & 2.37 (1.97-2.76) \\
6 & 14 & 2.40 (1.91-2.86) \\
8 & 20 & 1.87 (1.33-2.22) \\ \hline
\end{tabular}
\caption{Values of postshock electronic number density for a few regions (see Fig. \ref{hardness_ratio}) using
Eq. (\ref{ne_hat}) with $\theta_{\mathrm{s}} = 32 \arcmin$.
$D_{1}$ is the distance in units of 1 kpc.
The selected regions are those for which the 
thermal emission is detected for temperatures of $\sim 0.5-0.6$ keV.}
\label{tab_ne}
\end{table}

The postshock electronic number density
$\hat{n}_{\mathrm{e}} \equiv  \left( \int n_{\mathrm{e}}^2 \: dl / \int dl \right)^{1/2}$
is determined by (assuming that $n_{\mathrm{e}} = 1.21 \; n_{\mathrm{H}}$):
\begin{eqnarray}\label{ne_hat}
\hat{n}_{\mathrm{e}} & = & 8.21 \times 10^{-2} 
\left( \frac{ \int n_{\mathrm{e}} n_{\mathrm{H}} \: dl }{10^{16} \: \mathrm{cm}^{-5}}  \right)^{1/2} 
\left( \frac{D}{1 \: \mathrm{kpc}} \right)^{-1/2} \nonumber\\
& \times & \left( \frac{\theta_{\mathrm{s}}}{1 \arcmin} \right)^{-1/2} 
\left( 1- \left(\frac{\theta_{\mathrm{r}}}{\theta_{\mathrm{s}}}\right)^{2} \right)^{-1/4} \; \mathrm{cm}^{-3}
\end{eqnarray}
where the angular distance between the center of the SNR 
(taken as the position of the central point-like source 1WGA J1713.4--3949) and the
considered region is $\theta_{\mathrm{r}}$ while that of the blast-wave is $\theta_{\mathrm{s}}$.
Table \ref{tab_ne} shows that the average postshock electronic density is approximately
at the same level over the SNR ($\hat{n}_{\mathrm{e}} \sim 2.1^{+0.8}_{-0.9}\times 10^{-2}$ cm$^{-3}$ for $D = 1$ kpc), 
being a bit larger in the interior 
($\hat{n}_{\mathrm{e}} \sim 2.4 \pm 0.5 \times 10^{-2}$ cm$^{-3}$)
than close to the shock ($\hat{n}_{\mathrm{e}} \sim 1.9 \pm 0.5 \times 10^{-2}$ cm$^{-3}$).
In a Sedov model, the average postshock density is similar to the mean number density in the
ambient medium in the inner regions of the SNR.
Then, the postshock densities lead to a mean hydrogen number density $n_0$ of the ambient
preshock medium between $0.010$ and $0.024$ $D_{1}^{-1/2}$ cm$^{-3}$.

Nevertheless, because of the extremely low density in the ambient medium,
the assumption of equilibrium ionization state does not hold any more. 
Indeed, since the ionization age must be lower than
$\max(\hat{n}_{\mathrm{e}}) \: t_s = 9.0 \times 10^{8} 
\left( \frac{D}{1 \: \mathrm{kpc}} \right)^{1/2}
\left( \frac{t_s}{10^3 \: \mathrm{yr}} \right) \; \mathrm{s} \: \mathrm{cm}^{-3}$
where $t_s$ is the SNR age,
even a remnant of $10^5$ years old would be insufficient to reach the ionization equilibrium.
Consequently, we have applied a non-equilibrium model instead of the equilibrium model
while keeping the power-law model.
This has been done by relating the ionization age $\tau$ to the emission measure as described in Cassam-Chena\"{i} et al. (2004a, Sect. 6.3).
It is found that this model does not improve the fit obtained with the power-law model alone.
In the power-law plus non-equilibrium model,
the power-law characteristics are the same as those obtained for a single
power-law model (Table \ref{mo_wapo_mp}).
The temperature of the thermal gas is found to be around 3 keV but is poorly constrained.
Moreover, an upper limit on the ionization age of $4.5 \times 10^{8}$ s cm$^{-3}$ is obtained.
Assuming $t_s = 1600$ years (cf. Sect. \ref{energetics}), this leads to an upper limit of 
$2.6 \times 10^{-2} D_{1}^{-1/2}$ cm$^{-3}$ for the average postshock electronic density and
$0.02 \: D_{1}^{-1/2}$ cm$^{-3}$ for the mean hydrogen number density of the ambient
preshock medium ($D_{1}$ is the distance in units of 1 kpc).

Whatever the thermal model (equilibrium or non-equilibrium) added to the
nonthermal model, we derive similar mean hydrogen number densities $n_0$ of the ambient preshock medium.
In the following, we will adopt the upper limit $n_0 \leq 2 \times 10^{-2} \: D_{1}^{-1/2}$ cm$^{-3}$.

\begin{table*}[t]
\centering
\begin{tabular}{clllll}\hline \hline
One component model & $kT$ (keV) & $R$ $(D_{1}$ km) & $L$ $(10^{33} \: D^2_{1} \: \mathrm{erg} \: \mathrm{s}^{-1})$ & $\Gamma$ & $N_{\mathrm{H}}$ ($10^{22}$ cm$^{-2}$) \\ \hline \hline
Power law           &            &                                &          & 4.24     & 1.10 \\
$\chi^2 = 740$ (475 dof)  &      &                                &          & (4.18-4.30) & (1.07-1.13)\\ \hline
Blackbody           & 0.41       & 0.37                            &  0.49    &          & 0.35 \\
$\chi^2 = 622$ (475 dof) & (0.40-0.42) & (0.36-0.38)                &  (0.47-0.51) &       & (0.33-0.37) \\ \hline
Bremsstrahlung      & 0.94      &                                &          &            & 0.70 \\
$\chi^2 = 499$ (475 dof) & (0.92-0.96) &                          &          &            & (0.68-0.72) \\ \hline 
\end{tabular}
\caption{Best fit parameters for different one component spectral models for 1WGA J1713.4--3949 using the MOS and pn data all at once.
The errors are in the range $\Delta \chi^2 < 2.7$ (90\% confidence level) on one parameter. $D_{1}$ is the distance in units of 1 kpc.}
\label{XPSEC_1}
\end{table*}

\begin{table*}[t]
\centering
\begin{tabular}{clllll}\hline \hline
Two components model & $kT$ (keV)         & $R$ $(D_{1}$ km)  & $L$ $10^{33} \: (D^2_{1} \: \mathrm{erg} \: \mathrm{s}^{-1})$ & $\Gamma$   & $N_{\mathrm{H}}$ ($10^{22}$ cm$^{-2}$) \\ \hline \hline
Blackbody           & 0.40                &   0.31                          & 0.32           &            & 0.91 \\
$+$                 & (0.38-0.42)         &   (0.28-0.36)                    & (0.28-0.35)     &            & (0.85-0.97) \\
Power law           &                     &                                &               & 4.25         & \\
$\chi^2 = 503$ (473 dof)  &               &                                &               & (4.03-4.50)  &  \\ \hline 
Blackbody            & $kT_1$ = 0.57      & $R_1$ = 0.59                   & $L_1$ = 0.17  &            & 0.47 \\
$+$                  & (0.52-0.64)        & (0.51-0.69)                     &  (0.14-0.23)  &            & (0.44-0.51) \\
Blackbody            & $kT_2$ = 0.32      & $R_2$ = 0.11                    & $L_2$ = 0.44   &            &   \\ 
$\chi^2 = 490$ (473 dof) & (0.29-0.34)    & (0.07-0.16)                      & (0.40-0.48)     &            &   \\ \hline
\end{tabular}
\caption{Best fit parameters for different two components spectral models for 1WGA J1713.4--3949 using the MOS and pn data all at once.
The errors are in the range $\Delta \chi^2 < 2.7$ (90\% confidence level) on one parameter. $D_{1}$ is the distance in units of 1 kpc.}
\label{XPSEC_2}
\end{table*}

\begin{figure}[th]
\centering
\includegraphics[width=6cm,angle=-90]{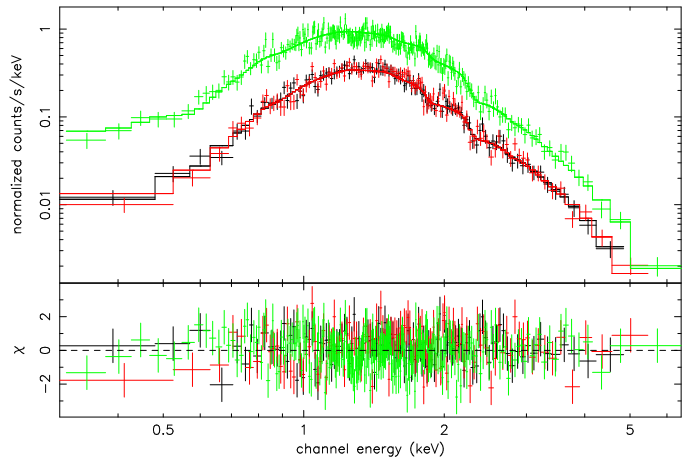}
\caption{EPIC spectra of the central source \mbox{1WGA J1713.4--3949} of the SNR RX J1713.7--3946 (MOS1 in black, MOS2 in red and pn in green).
Best-fit obtained with a double blackbody model.}
\label{NS_spectrum}
\end{figure}

\subsection{Central source \mbox{1WGA J1713.4--3949}\label{Neutron_Star}}

The nature of the point-like source \mbox{1WGA J1713.4--3949} located at the center of SNR RX J1713.7--3946
is not clearly established: a neutron star or a background extra-galactic source seem to be
possible interpretations (SL99). 
However, the neutron star hypothesis is the most attractive
because of the lack of radio emission and optical counterpart, 
and because the derived upper limit column density from the \textit{ASCA} X-ray spectra
is comparable with the total absorption through the Galaxy in this direction (SL99).

Recently, Lazendic et al. (2003) have shown that no pulsation and 
no long-term flux variability have been found using the \textit{ASCA}, \textit{RXTE}, \textit{Chandra} and \textit{XMM-Newton} data.
Besides, the radio pulsar PSR J1713-3945 
($\alpha_{\mathrm{J2000}}=$17h14m23.3s, $\delta_{\mathrm{J2000}}=$-39$^{\circ}$45$\arcmin$47.5$\arcsec$, $P=392$ ms, $D=4.3$ kpc) is not
associated with the X-ray point source \mbox{1WGA J1713.4--3949} nor with SNR RX J1713.7--3946.

Here, we intent to give better spectral constraints on the nature of this source \mbox{1WGA J1713.4--3949}
thanks to the high sensitivity of \textit{XMM-Newton}.
The spectral analysis of the source, as well as the knowledge of 
the absorbing column density map of the remnant (Sect. \ref{syn_emis}), 
must provide arguments in favor or against a potential association with
the SNR RX J1713.7--3946.

We extracted the source counts from a $25 \arcsec$ radius circle 
and the background was taken in an annulus of radii $R_{\mathrm{min}}=60 \arcsec$ and $R_{\mathrm{max}}=200 \arcsec$
around the point source center.
The background subtracted source count rates obtained are 
$0.410 \pm 0.006$, $0.413 \pm 0.006$ and $1.148 \pm 0.013$ counts s$^{-1}$ for MOS1, MOS2 and pn, respectively.
We extracted the spectrum of the central source in the 0.3-12.0 keV band. 
As shown in Fig. \ref{NS_spectrum}, it does not extend higher than 6.5 keV. 
No thermal spectral emission lines are found in the EPIC data, nor in the RGS data around 1 keV.

We first considered one component simple models, the results of which are summarized in Table \ref{XPSEC_1}.
Fitting the spectrum with a simple absorbed power law yields a large photon index $\Gamma = 4.24 \pm 0.06$.
Such a value is much steeper than what is generally found in other young neutron stars which are well known
for their non-thermal emission from relativistic particles accelerated in the pulsar magnetosphere 
(for instance, Willingale et al. 2001 give $\Gamma = 1.63 \pm 0.09$ for the Crab pulsar).
Moreover, the fit overestimates the spectrum at high energy resulting in a bad $\chi^2_{r}=1.58$ (for 475 dof).
The column density $N_{\mathrm{H}} = 1.10 \pm 0.03 \times 10^{22}$ cm$^{-2}$ 
inferred exceeds the values found in the SNR surrounding medium.

If the source is not an active pulsar but a neutron star, thermal radiation from the star surface can be observed.
A single blackbody model improves the fit ($\chi^2_{r}=1.31$ for 475 dof) 
but does not reproduce the hard emission.
It yields a high temperature of $kT = 0.40 \pm 0.2$ keV, a small radius 
$R = 0.37 \pm 0.01 \: D_{1}$ km of emitting area and a luminosity
of $L$ = $ 0.49 \pm 0.02 \times 10^{33} \: D^2_{1}$ $\mathrm{erg} \: \mathrm{s}^{-1}$ 
where $D_{1}$ is the distance in units of 1 kpc.
The value of $L$ is lower than predicted by standard cooling neutron star models 
for which the typical luminosity is $\sim 10^{34} \: \mathrm{erg} \: \mathrm{s}^{-1}$ for a 
neutron star of a few thousand years (e.g. Tsurata 1998).
Such a low luminosity could be explained by an accelerated cooling process.
However the small size of emitting area along with the high temperature means that the cooling
cannot take place from the entire neutron star surface.
Note that the temperature, radius and luminosity of the central point source \mbox{1WGA J1713.4--3949} 
are almost identical to what is found for the central point source in SNR RX J0852.0-4622 (also G266.2-1.2 or
``Vela Junior'') adopting a distance of 1 kpc (see Becker \& Aschenbach 2002, Kargaltsev et al. 2002).
The column density $N_{\mathrm{H}} = 0.35 \pm 0.02 \times 10^{22}$ cm$^{-2}$ 
is lower than the values found in the SNR.
We note that our range of temperatures is in the error bars given by SL99 whereas 
both our column density and radius are lower (in SL99,
$N_{\mathrm{H}} = 0.52^{+0.18}_{-0.16} \times 10^{22}$ cm$^{-2}$ and $R = 0.50^{+0.17}_{-0.11} \: D_{1} \: \mathrm{km}$).
The latter point should be related to the weak flux reduction.
Indeed, the unabsorbed flux in the $0.5-5.0  \: \mathrm{keV}$ energy band 
obtained with \textit{ASCA} ($5.3 \times 10^{-12} \: \mathrm{erg} \; \mathrm{cm}^{-2} \; \mathrm{s}^{-1}$, SL99)
is larger than the one obtained with \textit{XMM-Newton} 
($3.8 \times 10^{-12} \: \mathrm{erg} \; \mathrm{cm}^{-2} \; \mathrm{s}^{-1}$).

A simple thermal bremsstrahlung yields an excellent fit ($\chi^2_{r}=1.05$ for 475 dof) 
with a high temperature $kT = 0.94 \pm 0.2$ keV and an absorbing column density of
$N_{\mathrm{H}} = 0.70 \pm 0.02 \times 10^{22}$ cm$^{-2}$ compatible with the value derived
at the center of the SNR.

A more precise description of \mbox{1WGA J1713.4--3949} can be given by testing more 
elaborated models with two spectral components.
Table \ref{XPSEC_2} summarizes the results obtained.
To allow for different emission mechanisms, we first introduce a soft thermal component arising from
the neutron star surface with a blackbody model and a non-thermal component described by a power law due to
relativistic particles heating the polar caps of the neutron star.
The resulting fit is similar to that obtained with the thermal bremsstrahlung model.
The photon index $\Gamma = 4.24 \pm 0.06$ is steep in comparison to spectral index found in young neutron stars.
Both the luminosity and the temperature are a bit lower 
than what is expected for a cooling neutron star of a few thousand years (Tsurata 1998).
The column density $N_{\mathrm{H}} = 0.91 \pm 0.05 \times 10^{22}$ cm$^{-2}$ inferred exceeds the values found in the SNR center.

A two-component blackbody model allows to represent the non-uniformity
of the surface temperature distribution of the neutron star. 
The double blackbody is found to describe the data best ($\chi^2_{r}=1.04$ for 473 dof)
resulting in temperatures of $kT_1 = 0.57^{+0.07}_{-0.05}$ keV and $kT_2 = 0.32^{+0.02}_{-0.03}$ keV 
with $R_1 = 0.59^{+0.10}_{-0.09}\: D_{1}$ km  and $R_2 = 0.32^{+0.02}_{-0.03} \: D_{1}$ km for the radii of the emitting areas.
It suggests that the first thermal component is emitted from a small
fraction of the neutron star surface since the expected radius of a neutron star is $\sim 10-15$ km.
The second thermal component emission can be due to the heating of polar 
caps by relativistic particles which are accelerated in the magnetosphere or
to anisotropic heat conductivity in the neutron star crust (e.g. Becker \& Aschenbach 2002).
The fit yields a column density $N_{\mathrm{H}} = 0.47 \pm 0.03 \times 10^{22}$ cm$^{-2}$ 
that is compatible with the values found in the central part of the remnant.

These results are globally compatible with those of Lazendic et al. (2003).

\section{Discussion}\label{discussion}

\subsection{On the nature of the central source\label{SNII}}

To say whether the neutron star is associated with SNR RX J1713.7--3946 or not,
we must compare the results of our spectral analysis with the values found for
other types of compact objects found in other SNRs on one hand.
On the other hand, the absorbing column value derived from the best-fit model must be consistent 
with the absorbing column density found for the diffuse emission of the remnant in the central region.

The spectral properties (cf. Table \ref{XPSEC_2}) 
of the central point source \mbox{1WGA J1713.4--3949} are
very similar to those of the Compact Central Objects (CCOs) in ``Vela Junior'' (SNR G266.1-1.2 or RX J0852-4622),
in Cas A and in Puppis A (Becker \& Aschenbach 2002, Pavlov et al. 2002, Pavlov et al. 2004).
\footnote{The resemblance of RX J1713.7--3946 with ``Vela Junior'' is particularly striking
since both remnants have an identical type of central point source at the center
(no optical nor radio counterparts, absence of pulsations, same temperature, luminosity and surface emitting radius), 
are X-ray synchrotron dominated, have a radio emission weak and quite complex, 
live in a complex environment
and present an extremely low density of any thermally emitting material (Slane et al. 2001).}

In addition, the absorbing column $N_{\mathrm{H}} \sim 0.5 \times 10^{22}$ cm$^{-2}$
of the diffuse emission around central source \mbox{1WGA J1713.4--3949} 
(top panel of Fig. \ref{maps} and Table \ref{tab_NH_CO})
is consistent with the 
$0.47 \pm 0.03$ $10^{22}$ cm$^{-2}$ found for the two-component blackbody model (Table \ref{XPSEC_2}).
Note that this is also true considering the bremsstrahlung model 
which leads to a still acceptable 
$N_{\mathrm{H}} \sim 0.7 \times 10^{22}$ cm$^{-2}$ (Table \ref{XPSEC_1})
but is weakened adopting the blackbody+power-law model 
which yields a too high (almost a factor 2) column density $N_{\mathrm{H}} \sim 0.9 \times 10^{22}$ cm$^{-2}$ (Table \ref{XPSEC_2}).

As a conclusion, it is highly probable that the point source \mbox{1WGA J1713.4--3949}
is the compact relic of the supernova progenitor 
ranking RX J1713.7--3946 in the category of type II supernovae with strong particle acceleration.

\begin{figure}[t]
\centering
\begin{tabular}{c}
\includegraphics[width=10cm,bb=70 210 560 590]{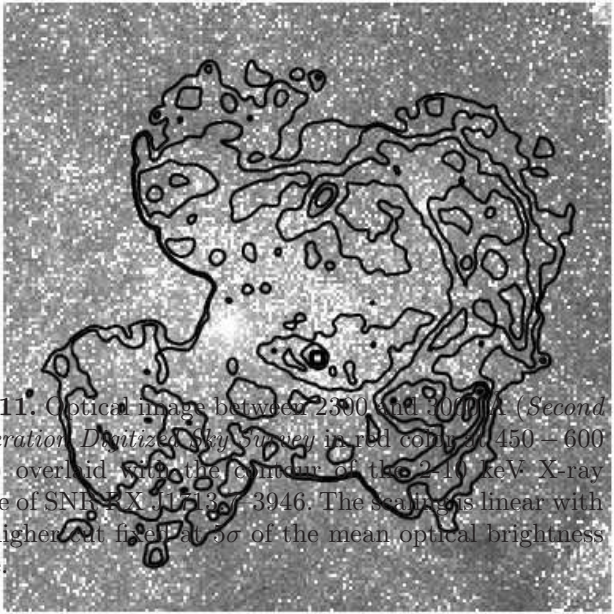} 
\end{tabular}
\caption{Optical image between 2300 and 3000 \AA \
(\textit{Second Generation Digitized Sky Survey} in red color 
at $450-600$ THz) 
overlaid with the contour of the 2-10 keV X-ray image of SNR RX J1713.7--3946.
The scaling is linear with the higher cut fixed at $5 \sigma$ of the mean optical brightness value.}
\label{im_optique}
\end{figure}

\begin{figure}[t]
\centering
\includegraphics[width=8cm]{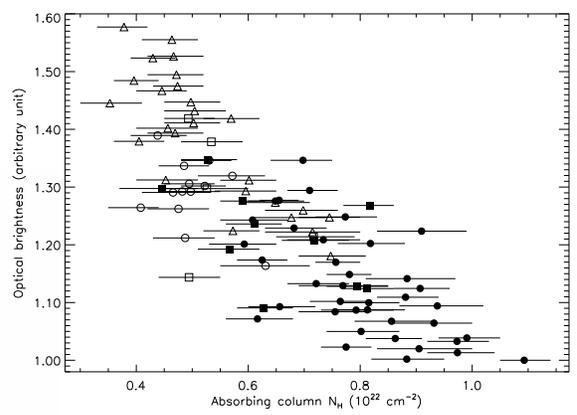}
\caption{Correlation plot between the absorbing column $N_{\mathrm{H}}$ 
(in units of $10^{22}$ cm$^{-2}$) and the normalized
optical brightness. 
To obtain this plot, the optical image shown in Fig. \ref{im_optique} 
has been cut following the absorbing column map grid.
Each point corresponds to the value found for a particular pixel of the mapping grid 
(Sect. \ref{mapping_method}) and
the error bars are given at a $90\%$ confidence level.
The symbols are defined as in Table \ref{duree_obs}.}
\label{correl_nh_dss}
\end{figure}

\subsection{Interaction with clouds and distance estimate}\label{interaction_with_clouds}
The mapping of the CO emission at $8\arcmin$ resolution 
shown in SL99 has revealed the presence of three dense and massive clouds
probably belonging to the same molecular complex of a Galactic arm at 6 kpc.
An enhanced intensity ratio of the two lowest rotational transition of the
CO molecule was found in one of these clouds located north 
to RX J1713.7--3946 suggesting that the SNR shock front is hitting this cloud.
Based on this interaction, it was concluded that the remnant 
should be at a distance of 6 kpc.

On the other hand, \textit{XMM-Newton} has revealed a positive correlation between the absorbing column 
density and the synchrotron X-ray brightness (see Fig. \ref{correl_nh_br}),
which strongly suggests that the brightest regions of the SNR interact with part of the
line-of-sight absorbing material (cf. Sect. \ref{syn_emis}).
In particular, this is the case of the bright southwestern region which exhibits the
strongest absorption in the line-of-sight.
At 6 kpc, however, the three dense and massive CO clouds are observed in the
north of the remnant; none is observed in the SW (SL99) where the highest absorption 
and a clear interaction are found with \textit{XMM-Newton}.

Independent information on the absorption in the line-of-sight of SNR RX J1713.7-3946
can be obtained from a map of integrated optical star light.
The optical star light is sensitive to the whole line-of-sight, 
but it should somehow reflect column density variations.
Figure \ref{im_optique} shows the \textit{red DSS} optical image.
The absorption morphology in the optical is quite different from 
the 6 kpc CO emission morphology presented in SL99 but similar to the X-ray absorption map.
The optical brightness appears to be high in the X-ray faint CE and
seems particularly weak in the X-ray bright SW region.
If we extract the optical brightness and $N_{\mathrm{H}}$ for 
each pixel of the mapping grid (Sect. \ref{mapping_method}),
we obtain a one dimension plot of these two quantities.
Figure \ref{correl_nh_dss} shows the very good inverse correlation obtained between our inferred column density 
and the optical brightness towards RX J1713.7-3946.
This is consistent with the low and high absorptions found
in the $N_{\mathrm{H}}$ map 
in the CE and SW, respectively (Fig. \ref{maps}, left panel).
This last result provides additional support to 
the reliability of the $N_{\mathrm{H}}$ distribution derived from the X-ray measurements.

The disagreement between the absorption morphology towards RX J1713.7-3946
and the molecular gas distribution at 6 kpc has an immediate consequence either
on the distance of the SNR or on the nature of the material interacting with the SNR.
If RX J1713.7-3946 is really located at 6 kpc, this implies that the remnant interacts with 
an interstellar component of different nature than molecular gas, as H\textsc{i} gas for instance.
In the other alternative, if RX J1713.7-3946 is really impacting molecular clouds, 
this implies that RX J1713.7-3946 is situated at a distance lower or larger than 6 kpc.
Because there is no evidence that the absorption toward the SNR is strongest in the north
contrary to what is observed in the CO map integrated up to 6 kpc (Bronfman et al. 1989, SL99),
it indicates that the remnant must be located at a distance lower than 6 kpc.

An indirect estimate of the density of the absorbing material located in the SW can be obtained 
from the local variations of column density in the interacting regions.
We assume that the angular size $\theta_{\mathrm{abs}}$ of the absorbing material is the same
along the line-of-sight and in projection on the sky.
If it is so, the density of the absorbing matter $n_{\mathrm{abs}}$ 
is given by:
\begin{equation}\label{n_abs}
n_{\mathrm{abs}} = 1.11 \; 10^{4} \; \left(\frac{D}{1 \; \mathrm{kpc}}\right)^{-1} 
\left(\frac{\theta_{\mathrm{abs}}}{1 \arcmin}\right)^{-1} 
\left(\frac{\Delta N_{\mathrm{H}}}{10^{22} \; \mathrm{cm}^{-2}}\right) \; \mathrm{cm}^{-3}
\end{equation}
where $D$ is the SNR distance and $\Delta N_{\mathrm{H}}$ is 
the variation of the absorbing column density between two regions separated by
angular size $\theta_{\mathrm{abs}}$.
Typically in the SW (see left panel of Fig. \ref{maps}), 
$\theta_{\mathrm{abs}} \sim 15\arcmin$ and 
$\Delta N_{\mathrm{H}} \simeq 0.4 \times 10^{22}$ cm$^{-2}$.
This gives 
$n_{\mathrm{abs}} \simeq 300 \; \left(\frac{D}{1 \; \mathrm{kpc}}\right)^{-1} \; \mathrm{cm}^{-3}$ 
which corresponds to $50 \; \mathrm{cm}^{-3}$ for a distance of the SNR of 6 kpc or 
$300  \; \mathrm{cm}^{-3}$ for 1 kpc,
a typical value for cold interstellar molecular clouds 
(e.g. Tatematsu et al. 1990, Wilner, Reynolds \& Moffett 1998).

The pending question is then to determine the nature 
and the distance of the matter interacting with RX J1713.7--3946.
Since the strongly absorbed regions are mainly found at the limit or around the SNR (particularly in the NW and SW)
and not inside, a possible scenario would be that the remnant
is within a wind blown bubble created by the precursor star, and that the
SN shock would be presently encountering the wall of such a bubble, as previously 
suggested (SL99, Ellison et al. 2001). 
However, the optical image shows that the thickness of
the regions of lowest optical brightness found at the edges of the remnant (Fig. \ref{im_optique})
is much larger than that expected from
a stellar wind shell. Then, the most plausible picture is
that of preexisting interstellar clouds, possibly pushed by the progenitor stellar wind, being struck by
the SNR shock front. This can be tested from CO and H\textsc{i} observations.

\begin{figure}[t]
\centering
\includegraphics[width=8cm,bb=54 360 450 720]{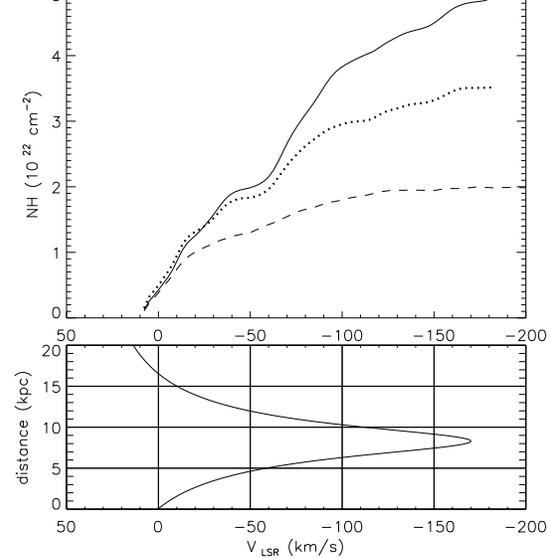}
\psfrag{NH}{$N_{\mathrm{H}}$}
\caption{\textit{Top panel}: The cumulative absorbing column density as
obtained from the atomic and molecular hydrogen, as a function of radial
velocity. The continuous line corresponds to the average NH estimated
towards the NW region, the dotted line corresponds to the SW region and
the dashed line to a line-of-sight directed to 1WGA J1713.4-3949.
\textit{Bottom panel}: The rotation curve towards l=347.5, b=-0.5 as obtained from
Fich et al.'s (1989) circular galactic rotation model.}
\label{nh_vLSR}
\end{figure}

\begin{table}[t]
\centering
\begin{tabular}{rrrrr} \hline \hline
\multicolumn{1}{c}{V$_{\mathrm{LSR}}$} & \multicolumn{1}{c}{$D$} & \multicolumn{3}{c}{$N_{\mathrm{H}}$ ($10^{22}$ cm$^{-2}$)}\\
 (km s$^{-1}$) & (kpc) & NW & SW & CE \\ \hline 
-0.4    &         0.14    &      0.43    &    0.50    &      0.38 \\
-3.0    &         0.58    &      0.52    &    0.60    &      0.48 \\
 \textbf{-5.6} & \textbf{0.98} & \textbf{0.62} & \textbf{0.71} & \textbf{0.56}  \\
 \textbf{-8.2} & \textbf{1.33} & \textbf{0.74} & \textbf{0.83} & \textbf{0.65} \\
 \textbf{-10.9} & \textbf{1.67} & \textbf{0.89} & \textbf{0.97} & \textbf{0.76} \\
-13.5   &         1.97    &      1.03    &   1.12     &     0.85 \\
-16.0   &         2.24    &      1.14    &   1.22     &     0.92 \\
-18.6   &         2.50    &      1.21    &    1.28    &      0.97 \\
-21.2   &        2.74     &     1.27     &   1.33     &     1.01 \\
-23.8   &         2.96    &      1.35    &    1.37    &      1.05 \\
-26.5   &        3.17     &     1.44     &   1.42     &     1.08 \\ \hline
\end{tabular}
\caption{Values of the accumulative absorbing column density
towards the NW, SW and CE as obtained from CO and H\textsc{i} observations
for several LSR radial velocities or equivalently for several distances.
The values of $N_{\mathrm{H}}$ which are in bold font are those compatible with the X-rays.}
\label{tab_NH_CO}
\end{table}

\begin{figure}[t]
\centering
\includegraphics[width=8cm]{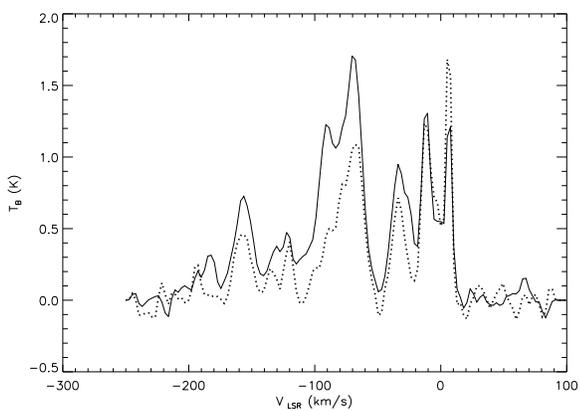}
\caption{Average CO spectra obtained towards the NW (continuous line)
and towards the SW (dotted line) from the CfA CO survey (Dame et al. 2001).
The SW emission is higher than the NW emission for LSR velocities between 0 and $-10$ km s$^{-1}$.}
\label{TB_vLSR}
\end{figure}

We have reanalyzed the behaviour of the absorbing material across the SNR 
RX J1713.7-3946 based on data from the CfA CO survey (Dame et al. 2001)
and from the IAR High Sensitivity H\textsc{i} survey (Arnal et al. 2000). The
angular resolution of the CO J:1-0 data is 8$\arcmin$.8, the grid 
sampling of 7$\arcmin$.5 and the velocity resolution of 1.3 km s$^{-1}$
(note that although Dame et al.'s 2001 new
survey improves the sampling interval to half a beamwidth, the region
containing the SNR RX J1713.7-3946 in the fourth Galactic quadrant, is
still based on the coarser sampling as published by Bronfman et al.
1989). For the H\textsc{i} data the angular resolution and the sampling are
30$\arcmin$ and the velocity resolution of 1.03 km s$^{-1}$.

Figure \ref{nh_vLSR} (top panel) displays the cumulative absorbing column density 
$N_{\mathrm{H}} = N$(H\textsc{i}) $+ \: 2 \times N$(H$_2$) as a
function of the \textit{Local Standard of Rest} (LSR) radial velocity, towards the NW (continuous
line), the SW (dotted line) and the center (dashed line). 
The plotted curves to the NW and SW were obtained 
from the average over $\sim 8\arcmin$ areas centered near
$\alpha_{\mathrm{J2000}}=$17h11m45s,
$\delta_{\mathrm{J2000}}=$-39$^{\circ}$32$\arcmin$45$\arcsec$ and 
near $\alpha_{\mathrm{J2000}}=$17h11m50s,
$\delta_{\mathrm{J2000}}=$-39$^{\circ}$57$\arcmin$00$\arcsec$, respectively.
The third $N_{\mathrm{H}}$ curve was calculated for a line-of-sight directed to
the bright central X-ray point-like source 1WGA J1713.4--3949. The
H$_2$ column density was calculated from the carbon monoxide
distribution by adopting the molecular mass calibration ratio 
X= ($1.8 \pm 0.3) \times 10^{20}$cm$^{-2}$  (Dame et al. 2001). 
The behaviour of Fig. \ref{nh_vLSR} (top panel) is mainly 
defined by the CO profile
because of the better angular resolution of these data as
compared with the used H\textsc{i} survey and its dominant contribution.
The galactic rotation curve for $l= 347.5^\circ, b=-0.5^\circ$ (from Fich
et al.'s 1989 circular galactic rotation model) is plotted in
Figure \ref{nh_vLSR} (bottom panel) for comparison.
Table \ref{tab_NH_CO} gives the value of the cumulative absorbing column density
towards the NW, SW and center for several LSR radial velocities or 
equivalently for several distances (only the closer distance is provided).

The line-of-sight towards $l= 347.5^\circ$ basically crosses three
galactic arms: Sagittarius, at an approximate distance of 1.4 kpc
(V$_{\mathrm{LSR}} \sim -10$ km s$^{-1}$), Scutum--Crux at about 3.7 kpc
(V$_{\mathrm{LSR}} \sim -34$ km s$^{-1}$), and Norma at about 5.3 kpc
(V$_{\mathrm{LSR}} \sim -66$ km s$^{-1}$) (Georgelin \& Georgelin 1976).
Figure \ref{TB_vLSR} displays two average spectra obtained towards the NW (continuous line)
and towards the SW (dotted line), where, in addition to the local gas,
the three important galactic components are evident.

From the X-ray absorption map (see Fig. \ref{maps}, left panel), the
absorbing column density was found to be higher in the SW 
($N_{\mathrm{H}} \sim 0.8-1.1 \times  10^{22}$ cm$^{-2}$) than in the NW
($N_{\mathrm{H}} \sim 0.6-0.7 \times 10^{22}$ cm$^{-2}$).
The $N_{\mathrm{H}}$ curve obtained to the
SW from CO and H\textsc{i} observations, lies above the curve
obtained for the NW region only for velocities below
 V$_{\mathrm LSR} \sim -23$ km s$^{-1}$ as shown in Fig. \ref{nh_vLSR} (top panel).
Although the limited angular 
resolution of the available H\textsc{i} data ``dilutes'' 
small spatial variations 
and reduces the $\Delta N_{\mathrm{H}}$ between the two regions,
this sets an upper limit on the SNR distance at $\sim$ 2.9 kpc using
the galactic rotation curve (Fig. \ref{nh_vLSR}, bottom panel).

The absorbing column densities derived from the \textit{XMM-Newton} data
are found to be $0.4-0.6 \times 10^{22}$ cm $^{-2}$ towards 
1WGA J1713.4--3949, $0.6-0.7 \times 10^{22}$ cm$^{-2}$  towards the NW, 
and $0.8-1.1 \times 10^{22}$ cm$^{-2}$ towards the SW (Sect. \ref{syn_emis}).
These values match very well those found from the CO and H\textsc{i} observations
if the LSR radial velocity range is between $\sim -6$ and $\sim -11$ km s$^{-1}$ as
indicated on Table \ref{tab_NH_CO}.
Then, it reduces the SNR distance range to $\sim 0.9-1.7$ kpc.

Besides, the only negative LSR velocities for which the CO emission
is clearly higher in the SW than in NW are found between 0 and $-10$ km s$^{-1}$
as revealed by the spectra displayed in Fig. \ref{TB_vLSR}.
Indeed, it can be noticed
the presence of a ``shoulder'' of higher emission obtained to the SW (dotted line),
as compared to the NW spectra (solid line).
This constitutes an evidence for the presence of
molecular gas in the SW direction.
This feature is consistent with the
molecular cloud recently detected by Fukui et al. (2003) in the SW 
(between $-3$ and $-11$ km s$^{-1}$) on the basis of 2.6 mm 
CO J:1-0 observations carried out with \textit{NANTEN} telescope.
The range of LSR velocities corresponding to the upper limit distance of $1.6$ kpc,
using the circular galactic rotation model (Fich et al. 1989),
is complety compatible with the allowed distance range of $\sim 0.9-1.7$ kpc.

A search for ground-state satellite line of the hydroxyl molecule at
1720.53 MHz was carried out towards two spots in G347.3-0.5 where
both, X-rays and CO observations, suggested strong shock/molecular cloud
interaction. The positive detection of OH (1720 MHz)
masers  permits: to confirm the existence of such
interaction and to constrain the physical parameters of the
shock-excited gas on the basis of the very specific conditions required
to excite the 1720 MHz OH masers.
 
The observations were carried out with the Very Large Array
of NRAO\footnote{The National Radio Astronomy Observatory is a
facility of the National Science Foundation, operated under
cooperative agreement by Associated Universities, Inc.} in the
C-configuration for 2 hours on March 11, 2004. Two 22\arcmin~ fields
centered at (17h11m45s, $-39^\circ 57\arcmin 30\arcsec$) and
(17h11m45s, $-39^\circ 57\arcmin 30\arcsec$) were observed.
A 256-channels spectrometer centered around -30 km s$^{-1}$, with a channel
width of 3.05 kHz/channel ($\sim$ 0.5  km s$^{-1}$ at 1720 MHz) and a
velocity range $\Delta$v $\sim 134$ km s$^{-1}$, was used.  The data
were cleaned and calibrated using AIPS routine programs. The
synthesized beam was $40\arcsec \times 10\arcsec$ and the noise level
of about 50 $\mu$Jy/beam per channel.
 
The results of the search for OH (1720 MHz) masers were negative. This
fact is probably due either to the fact that the shocked molecular clouds
have a volume density smaller than required to excite the masers, or
the shock front was not of a non-dissociative C-type (Wardle 1999).

The X-ray analysis of SNR RX J1713.7--3946 with \textit{XMM-Newton} has revealed a strong absorption in the SW,
even stronger than in the NW. 
The absorption derived from the optical star brightness is in agreement with the X-ray derived map.
This high absorption found in the SW was unexpected with regard to the morphology of the CO distribution at 6 kpc.
Moreover, the positive correlation between X-ray absorption and the X-ray brightness is indicating that 
the shock front of RX J1713.7--3946 is impacting the molecular clouds responsible for the absorption.
The CO and H\textsc{i} observations show that the inferred cumulative absorbing column densities 
are in excellent agreement with the X-ray findings in different places of the remnant
provided that the SNR lies at a distance of $1.3 \pm 0.4$ kpc, probably in the Sagittarius galactic arm.
Other indications of an interaction between the SNR blast-wave and molecular clouds
are provided by the existence of an excess of fast-moving H\textsc{i} gas (Koo et al. 2004)
and possibly, the existence of a broad CO profile in LSR velocity found in a 
coincident bright X-ray feature in the SW at 1 kpc (Fukui et al. 2003).
Based on the absorption morphology and the probable association between the SNR and the molecular clouds,
the distance of SNR RX J1713.7--3946 is revised to $1.3 \pm 0.4$ kpc instead of 
the commonly accepted value of 6 kpc.

\subsection{Particle acceleration}\label{acceleration}

The asymmetry in the synchrotron X-ray emission morphology
is the result of the asymmetry in the ambient density around the remnant.
In the strongly absorbed bright SW rim, we have established that 
the shock front of RX J1713.7--3946 is interacting with molecular clouds
as suggested by the positive correlation between X-ray absorption and X-ray brightness.
In the weakly absorbed faint SE region, the same correlation suggests that
the shock front of RX J1713.7--3946 is propagating into a tenuous ambient medium 
which is consistent with the larger extent of the SNR than in the SW.
In Sect. \ref{syn_emis}, we have looked at the spatial modifications (in the SW and SE) 
of particle acceleration at the shock in RX J1713.7--3946 through the photon spectral index changes.
The regions where the shock impacts molecular clouds have a steeper spectrum
($\Gamma \sim 2.3-2.7$) than those where the shock propagates
into a low density medium ($\Gamma \sim 1.9-2.2$).

Qualitatively, we can interpret such modifications in the synchrotron spectrum 
by considering a modification of the acceleration in the interaction region with dense molecular clouds.
To investigate that effect, we have used a simple model of nonlinear diffusive shock 
acceleration (Berezhko \& Ellison 1999, Ellison et al. 2000) 
varying the upstream density $n_0$, the shock speed $V_{\mathrm{sk}}$
and the magnetic field $B_0$ for an injection (\emph{i.e.}
fraction of gas particles which are accelerated) set to $10^{-3}$.
As the SNR blast-wave runs into a high density medium such as molecular clouds, 
its velocity is expected to diminish according to the ram pressure conservation ($n_0 \: V_{\mathrm{sk}}^2$ constant) 
and the upstream magnetic field is expected to increase following the square root of the density ($B_0 \: n_0^{-1/2}$ constant).
The result of the density increase is to lead to a steepening of the
synchrotron spectrum but no increase in brightness. Therefore, this cannot explain
the observed difference in X-ray brightness between the SE and the SW (for the same magnetic field orientation).
However, if the shock is preferentially tangential in the region of high density but normal in the region of lower density, 
the X-ray brightness can be strongly enhanced due to the compression of the downstream magnetic field on condition that
the density contrast between the two regions is weak (not more than a factor 4).
Then, it is possible to explain stronger X-ray brightness and steeper synchrotron spectrum in the SW than in the SE
if the magnetic field is tangential in the SW but normal in the SE and if the acceleration process 
occured before the shock entered the dense molecular clouds.

A modeling of these effects is beyond the scope of this paper.
Also, further smaller scale spectral studies with the \textit{Chandra} satellite would help to
investigate the origin of the steepening of the spectrum in the SW.

\subsection{SNR age and energetics}\label{energetics}

The study of the thermal emission with \textit{XMM-Newton} has lead to an upper limit on the density
in the ambient medium of $n_0 \leq 2 \times 10^{-2}  \: D_{1}^{-1/2}$ cm$^{-3}$ 
where $D_{1}$ is the distance in units of 1 kpc (cf. Sect. \ref{thermal_emission}).
If the density derived from the emission measure is indeed representative of the
density in the ambient medium (which is not the case in the radiative phase),
an estimate of the swept-up mass is given by:
\begin{equation}\label{Msw}
M_{\mathrm{sw}} \sim 2.96 \times 10^{-3} 
\left( \frac{D}{1 \: \mathrm{kpc}} \right)^{3}
\left( \frac{\theta_{\mathrm{s}}}{1 \arcmin} \right)^{3} 
\left( \frac{ n_{0} }{ 1 \: \mathrm{cm}^{-3} } \right) \; \mathrm{M}_{\odot}
\end{equation}
where $\theta_{\mathrm{s}}$ is the angular size of the blast-wave radius.
This leads to an upper limit on the swept-up mass of $1 \; \mathrm{M}_{\odot}$ for a distance of 1 kpc and
$\theta_{\mathrm{s}} = 32\arcmin$.
If RX J1713.7--3946 is the remnant of a type II explosion, 
the initial mass of the progenitor is expected to be above $5-8 \; \mathrm{M}_{\odot}$ (Kennicutt 1984)
and then the mass of the ejected matter of a few solar masses. 
This is well above the derived swept-up mass leading to the conclusion that 
RX J1713.7--3946 must be in the free expansion phase.

The hypothesis proposed by Wang et al. (1997),
based on historical records, that RX J1713.7--3946 is the 
remnant of the supernova that exploded in AD 393 would imply an age of $t_s = 1600$ years.
Assuming self-similar solutions (Chevalier 1982), we can estimate the shock velocity
given by
$v_s = 2.85 \times 10^{2} \: (n-3)/(n-s) \left( D / 1 \: \mathrm{kpc} \right)
\left( \theta_{\mathrm{s}} / 1 \arcmin \right)
\left( t_s / 10^3 \: \mathrm{yr} \right)^{-1} \; \mathrm{km} \: \mathrm{s}^{-1}$ 
where $n$ and $s$ are the power-law indexes of the initial outer density profile in the ejecta and
of the initial ambient medium density profile, respectively.
$n \sim 9-12$ is typical of SN II, 
$s=0$ corresponds to an homogeneous ambient medium, $s=2$ to an ambient medium modified by the progenitor wind.
For $n=9$, $v_s \simeq 3.8 \times 10^{3} \; \mathrm{km} \: \mathrm{s}^{-1}$ if $s=0$
and $v_s \simeq 4.9 \times 10^{3} \; \mathrm{km} \: \mathrm{s}^{-1}$ if $s=2$.

The determination of the kinetic energy of the explosion provides a strong argument for or against
the hypothesis that RX J1713.7--3946 is the remnant of the AD 393 stellar explosion.
In the framework of spherical self-similar model (Chevalier 1982), 
we can estimate the kinetic energy of the explosion as a function of the SNR age 
for different progenitor masses and ambient medium densities.
The energy curves depend on three parameters among which $n$ and $s$ for a given progenitor mass (Decourchelle \& Ballet 1994).
The power-law index $n$ of the ejecta density profile was fixed to 9.
The power-law index $s$ of the ambient density profile was fixed to 0 to simulate a flat ambient medium which seems
to be the situation in the SE.
The kinetic energy of the explosion is then determined by the relation:
\begin{eqnarray}
E_{\mathrm{SN}} & = & \frac{3}{10} (n-5)^{-1} \: (n-3)^{(n-5)/(n-3)}  
\left( \frac{4 \: \pi \:n \: \rho_0}{ 3 \: A} \right)^{ 2/(n-3) } \nonumber\\
  & \times & \left( \epsilon \: r_s \right)^{ 2(n-s)/(n-3) } M_{\mathrm{ej}}^{ (n-5)/(n-3) } \; t_s^{-2}
\end{eqnarray}
where $r_s$ is the shock radius set to $10$ pc, $\rho_0 = 1.4 \: m_{\mathrm{H}} \: n_0$ and $t_s$ the SNR age.
The parameters $A$ and $\epsilon$ are determined by solving the self-similar hydrodynamical equations (Chevalier 1982).

\begin{figure}[th]
\centering
\includegraphics[width=8cm]{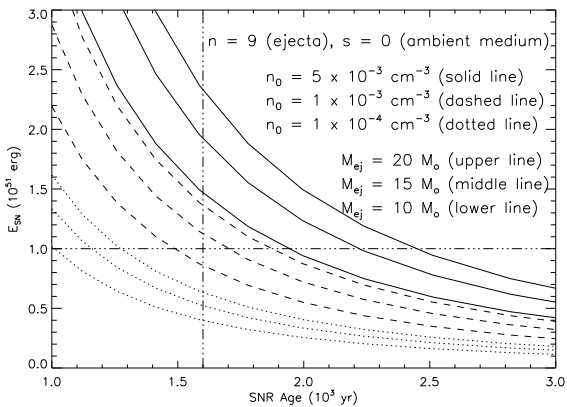}
\caption{Kinetic energy of the explosion as a function of the SNR age for several solar masses of
the supernova ejecta ($M_{\mathrm{ej}} = 10, 15$ and $20 \: \mathrm{M}_{\odot}$) and different ambient medium densities
($n_0 = 5 \times 10^{-3}, 1 \times 10^{-3}$ and $5 \times 10^{-4}$ cm$^{-3}$).}
\label{E51_age}
\end{figure}

Energy curves are illustrated in Fig. \ref{E51_age}
for supernova ejecta masses of $M_{\mathrm{ej}} = 10, 15$ and $20 \: \mathrm{M}_{\odot}$ representative of type II supernova
and for different ambient medium densities between $1 \times 10^{-4}$ and $5 \times 10^{-3}$ cm$^{-3}$.
If we adopt the standard $10^{51}$ erg as the energy of the explosion,
it shows that an ambient medium density of $10^{-3}$ cm$^{-3}$ is appropriate 
for ejecta masses between $10 \: \mathrm{M}_{\odot}$ and $15 \: \mathrm{M}_{\odot}$ if we consider a remnant of 1600 years old.
At this age, higher density of the ambient medium ($5 \times 10^{-3}$ cm$^{-3}$)
leads to a higher kinetic energy of the explosion, still possible, whereas
below $10^{-4}$ cm$^{-3}$, the kinetic explosion energy begins to be too weak

If we consider the effect of efficient cosmic-ray acceleration on the structure of the blast-wave
as described in Chevalier (1983), energy curves move a bit higher and
would only very slightly widen the range of allowed ambient medium densities.
Hence, these results based on energy argumentation are consistent with the 
picture of a young SNR expanding in a low density medium.

\subsection{Evolutionary stage}\label{evol_stage}

Although the star formation process is complex, there is observational
evidence that massive stars are born in the densest parts of molecular clouds.
The presence of molecular clouds around the northern and southwestern parts of
RX J1713.7--3946 strongly suggests that its progenitor star is indeed born in such conditions.
Evolved SNRs interacting with molecular clouds are expected to be efficient electron accelerators 
(Bykov et al. 1999) as observed in RX J1713.7--3946.
Nevertheless, these molecular clouds appear to be located at the periphery of the SNR at
$\sim 7$ pc away from the center, assuming a distance of 1 kpc (here and hereinafter).
Massive stars are effectively known for initiating and maintaining powerful 
stellar winds during their lives so that they can move the molecular clouds
from the vicinity of the progenitor star.
These winds may have homogeneized the immediate environment of the progenitor star,
explaining the extremely low mean density inside the SNR (McKee et al. 1984, Chevalier 1999).

In the standard picture of a stellar wind interacting with the interstellar medium,
a four-zone structure (Castor et al. 1975, Weaver et al. 1977) settles very rapidly, consisting of:
the innermost hypersonic stellar wind characterized by a density profile 
$\rho \propto r^{-2}$,
a hot shocked stellar wind region,
a shell of shocked interstellar medium,
and the ambient interstellar gas.
After a few thousand years,
the swept-up shocked interstellar gas collapses into a thin cold ($T \sim 10^4$ K) 
shell strongly affected by radiative losses while
the low-density ($< 1$ cm$^{-3}$) shocked stellar wind region occupying most of the bubble volume,
remains very hot ($T > 10^6$ K).

A possible interpretation of our observation of RX J1713.7--3946 is that the SN event occured in a wind-blown bubble.
In the SE, the shock front of the remnant is still propagating into the 
low-density hot shocked wind ($n_{0} \leq 0.02$ cm$^{-3}$).
In the SW, the blast-wave of RX J1713.7--3946
has already progagated into the shocked stellar wind and is impacting the cold and dense
($\sim 300$ cm$^{-3}$, see Eq. (\ref{n_abs})) shell of shocked ambient material.
The remnant could be in the free expansion phase in the SE whereas in the radiative phase in the SW.

In this scenario, the size of the bubble provides an estimate of the strength of the wind and hence
of the progenitor mass. 
From Fig. \ref{X_ray_images}, the position of the bright X-ray rims in the NW and SW
($r_{\mathrm{b}} \sim 6-9$ pc, taking into account the projection effects)
provides an estimate of the radius of the bubble $r_{\mathrm{b}}$, assuming that the molecular clouds have been
removed from the vicinity of the star by the progenitor stellar wind.

\begin{table}[t]
\centering
\begin{tabular}{lllllr}
\hline \hline
\multicolumn{1}{c}{$M_{\star}$} & \multicolumn{1}{c}{Spectral} & \multicolumn{1}{c}{$\dot{M}_{\mathrm{loss}}$} & \multicolumn{1}{c}{$L_{\mathrm{w}}$} & \multicolumn{1}{c}{$t_{\mathrm{ms}}$} & \multicolumn{1}{c}{$r_{\mathrm{b}}$} \\
\multicolumn{1}{c}{($\mathrm{M}_{\odot}$)} & \multicolumn{1}{c}{type} & \multicolumn{1}{c}{($\mathrm{M}_{\odot} \: \mathrm{yr}^{-1}$)} & \multicolumn{1}{c}{(erg s$^{-1}$)} & \multicolumn{1}{c}{(yr)} & \multicolumn{1}{c}{(pc)}  \\ \hline
12 & B1 V & $6 \times 10^{-9}$ & $9.3 \times 10^{32}$ & $1.3 \times 10^{7}$ & 5.3 \\
16 & B0 V & $6 \times 10^{-8}$ & $9.3 \times 10^{33}$ & $0.9 \times 10^{7}$ & 10.0 \\ \hline
\end{tabular}
\caption{Parameters for main sequence evolution for different progenitor masses $M_{\star}$ (from Chevalier 1999 and references therein).
$\dot{M}_{\mathrm{loss}}$ is the mass loss rate,
$L_{\mathrm{w}}$ is the mechanical wind luminosity,
$t_{\mathrm{ms}}$ is the main-sequence age and $r_{\mathrm{b}}$ is the
maximum size of the wind bubble.}
\label{MSstars}
\end{table}

Then, an estimate of the mass of RX J1713.7--3946's progenitor star can be obtained by 
directly comparing its wind-bubble size where the molecular clouds are located
with those estimated from observations of stars of different masses.
Table \ref{MSstars} shows two cases for which the bubble radii (5.3 and 10.0 pc) 
are close to the $6-9$ pc values estimated in the NW and SW for RX J1713.7--3946.
This implies that the mass of RX J1713.7--3946's progenitor must be between 
$12$ and $16 \: \mathrm{M}_{\odot}$ in the main sequence evolution phase, which
corresponds to a B-star whose lifetime on the main sequence is about $10^7$ years.
In Table \ref{MSstars}, the mechanical wind luminosity $L_{\mathrm{w}} \equiv \dot{M}_{\mathrm{loss}} v_{\mathrm{w}}^2 / 2$ 
is computed assuming an appropriate wind velocity $v_{\mathrm{w}} = 700$ km s$^{-1}$ as chosen in Chevalier (1999).
The maximum size of the wind bubble is given by (in pc):
\begin{equation} 
r_{\mathrm{b}} = 10.7 
\left( \frac{ L_{\mathrm{w}} }{ 10^{34} \: \mathrm{erg} \: \mathrm{s}^{-1} }\right)^{1/3} 
\left( \frac{ p/k }{ 10^{5} \: \mathrm{K} \: \mathrm{cm}^{-3} }\right)^{-1/3}
\left( \frac{ t_{\mathrm{ms}} }{ 10^{7} \: \mathrm{yr} }\right)^{1/3}
\end{equation}
assuming that the 
surrounding interclump medium exerts a pressure $p/k = 10^5 \: \mathrm{K} \: \mathrm{cm}^{-3}$, typical of cloud (Chevalier 1999).

From the characteristics of B-stars, we can estimate the mean density in the wind
before the supernova event
and compare it to the density derived from the thermal X-rays (see Sect. \ref{thermal_emission}).
The mass lost during the lifetime of the progenitor diluted in the volume of the bubble gives 
an estimate of the mean number density $n_{\mathrm{b}}$ of the wind-bubble (in cm$^{-3}$):
\begin{equation}\label{n_bubble}
n_{\mathrm{b}} = 6.9 \times 10^{-4} 
\left( \frac{\dot{M}_{\mathrm{loss}}}{10^{-8} \: \mathrm{M}_{\odot} \: \mathrm{yr}^{-1}} \right)
\left( \frac{t_{\mathrm{ms}}}{10^7 \: \mathrm{yr}} \right)
\left( \frac{r_{\mathrm{b}}}{10 \: \mathrm{pc}} \right)^{-3}
\end{equation}
For $M_{\star} = 12$ and $16 \: \mathrm{M}_{\odot}$ (cf. Table \ref{MSstars}), we obtain a mean density of 
$3.6 \times 10^{-3}$ and $3.7 \times 10^{-3}$ cm$^{-3}$, respectively. 
This is consistent with the $\sim 2 \times 10^{-2}$ cm$^{-3}$ upper limit found for the thermal emission in the X-rays.
Note that the mass lost in the red giant phase was not taken into account because 
it produces a local excess in density only in the vicinity of the progenitor star.

The proposed scenario in which the SNR blast wave expands into a low-density shocked wind medium
in the SE and strikes molecular clouds in the SW
appears to be a possible physical picture since
the density derived from the thermal X-ray emission measures can be found to be consistent
with the expected density in the pre-SN shocked stellar wind.
However, it is hard to understand why the shell of molecular clouds shocked by the SNR blast-wave
does not produce copious thermal X-ray emission since the X-ray brightness is proportional to the square of the density. 
The regions of enhanced X-ray emission as observed in the SW are
those where there is a strong increase of the synchrotron X-ray emission
(correlated with an increased of absorbing column density),
but where there is no indication of any increase of the thermal X-ray emission.

However, if the SW rim of RX J1713.7--3946 is in the radiative phase, then we do not expect thermal X-ray
emission from the shocked interstellar clouds.
It would also explain why the thermal X-ray density does not follow the variations of the ambient density.
A high density of $300$ cm$^{-3}$ and a temperature of $10^5$ K in the shocked interstellar clouds
yield a cooling time of $\sim 25$ years.

However, in SNRs where the shock is radiative, as observed in the Cygnus Loop,
bright optical filaments are generally observed (Levenson et al. 1996).
The optical emission is attributed to the cooling of the gas behind the shock front.
This cooling gas emits many optical lines including hydrogen Balmer lines with a temperature of $\sim 10^4$ K.
If RX J1713.7--3946 is indeed in the radiative phase
where the blast wave has reached the cavity wall and is interacting with dense matter, such optical filaments
might be visible, although the absorption along the line-of-sight is much stronger in  
RX J1713.7--3946 ($N_{\mathrm{H}} \sim 10^{22}$ cm$^{-2}$) than in
the Cygnus Loop SNR ($N_{\mathrm{H}} \sim 4 \times 10^{20}$ cm$^{-2}$).

Finally, we could expect an enhancement of the thermal X-ray emission
in the shocked stellar wind just behind the shocked shell. Indeed,
a blast wave impacting on a dense cloud gives rise to both a reflected shock propagating back
and a transmitted shock propagating into the cloud (McKee \& Cowie 1975).
The shocks that are reflected off molecular clouds surfaces 
are expected to propagate back into the hot,
previously shocked gas, further heating and compressing it (e.g., Levenson et al. 1996).
Then, enhanced X-ray thermal emission is expected (Hester \& Cox 1986).
However, the density in the shocked gas is still small so that its contribution might be negligeable.

\section{Conclusion}\label{conclusion}
The \textit{XMM-Newton} observation has yielded several new results on SNR RX J1713.7--3946.
Observational facts and their astrophysical interpretations are summarized here:
\begin{itemize}
\item The X-ray bright central point source \mbox{1WGA J1713.4--3949} detected at the center of SNR RX J1713.7--3946
shows spectral properties very similar to those of the Compact Central Objects  found in SNRs (e.g., Vela Junior) and which are
consistent with the absorbing column density of the central diffuse X-ray emission arising from the SNR.
It is highly probable that the point source \mbox{1WGA J1713.4--3949}
is the compact relic of RX J1713.7--3946's supernova progenitor.
\item The X-ray mapping of the absorbing column density has revealed strong variations 
($0.4 \times 10^{22}$ cm$^{-2} \leq N_{\mathrm{H}} \leq 1.1 \times 10^{22}$ cm$^{-2}$)
and, particularly, an unexpected strong absorption in the southwest.
The column density variations are well reflected by the extinction in the map of integrated optical star light.
The strong positive correlation between the X-ray absorption and the X-ray brightness along the western rims
suggests that the shock front of RX J1713.7--3946 is impacting molecular clouds there.
\item CO and H\textsc{i} observations 
show that the inferred cumulative absorbing column densities 
are in excellent agreement with the X-ray measurements in different places of the remnant only
if the SNR is placed at a distance of $1.3 \pm 0.4$ kpc, probably in the Sagittarius galactic arm, 
and not at 6 kpc as previously claimed.
An excess in the CO emission found in the southwest at $\sim 1$ kpc strongly suggests 
that molecular clouds produce the enhancement in absorption.
A search for OH masers in the southwestern region has been unsuccessful, possibly due to the low density of the clouds.
\item The X-ray mapping of the photon index has revealed strong variations ($1.8 \leq \Gamma \leq 2.6$).
The spectrum is steep in the faint central regions and flat at the 
presumed shock locations, particularly in the southeast.
However, the regions where the shock strikes molecular clouds have a steeper spectrum
than those where the shock propagates into a low density medium.
\item The search for the thermal emission in RX J1713.7--3946 has been still unsuccessful leading
to a number density upper limit of $2 \times 10^{-2}$ cm$^{-3}$ in the ambient medium.
This low density in the ambient medium yields to a reasonable kinetic energy of the 
explosion provided that the remnant is less than a few thousand years old.
It can be reconciled with the high density in molecular clouds if the remnant is in the radiative phase 
where the SNR shock encounters a dense ambient medium whereas it is in the free expansion phase elsewhere.
\item RX J1713.7--3946's progenitor mass is estimated to lie between 12 and $16 \: \mathrm{M}_{\odot}$
based on a scenario involving the effect of stellar wind of the progenitor star.
\end{itemize}

\begin{acknowledgements}
We acknowledge W. M. Goss for performing the OH VLA observations.
G.C.-C. is deeply indebted to Isabelle Grenier for many helpful discussions on H\textsc{i} and CO observations
and also on neutron stars.
The results presented here are based on observations obtained with \textit{XMM-Newton},
an ESA science mission with instruments and contributions 
directly funded by ESA Member States and the USA (NASA).
Part of the project was supported by the CONICET (Argentina) - CNRS (France) cooperative
science program No. 10038.
\end{acknowledgements}




\begin{thebibliography}{}

\bibitem[Arnal et al. 2000]{arnal}
Arnal, E. M., Bajaja, E., Larrarte, J. J., Morras, R., Poppel, W., 2000, A\&AS, 142, 35

\bibitem[Arnaud et al. 2001]{arnaud}
Arnaud, M., Neumann, D. M., Aghanim, N., et al. 2001, A\&A, 365, L80

\bibitem[Arnaud et al. 2002]{arnaud2}
Arnaud, M., Majerowicz, S., Lumb, D., et al. 2002, A\&A, 390, 27

\bibitem[Arnaud 1996]{arnaud_2}
Arnaud, K. A. 1996, ASP Conf. Ser. 101, Astronomical Data Analysis Sofware and Systems V, ed. G. Jacoby, \& J. Barnes, 17.

\bibitem[Becker \& Aschenbach 2002]{becker}
Becker, W, \& Aschenbach, B. 2002, Proceedings of the 270, WE-Heraeus Seminar on Neutrons Stars, Pulsars and Supernova Remnants

\bibitem[Berezhko \& Ellsion 1999]{ellison}
Berezhko, E. G., \& Ellison, D. C. 1999, ApJ, 526, 385

\bibitem[Bronfman et al. 1989]{bronfman}
Bronfman, L., Alvarez, H., Cohen, R. S., Thaddeus, P., 1989, ApJS, 71, 481

\bibitem[Butt et al. 2001]{butt}
Butt, Y. M., Torres, D. F., Combi, J. A., Dame, T., \& Romero, G. E. 2001, 
ApJ, 562, L167
\bibitem[Butt et al. 2002]{butt2}
Butt, Y. M., Torres, D. F., Romero, G. E., Dame, T. M., \& Combi, J. A. 2002, Nature, 418, 499

\bibitem[Bykov et al. 2000]{bykov}
Bykov, A. M., Chevalier, R. A., Ellison, D. C., \& Uvarov, Y. A. 2000, ApJ, 538, 203

\bibitem[Cassam-Chena\"{i} et al. 2004a]{gcc1}
Cassam-Chena\"{i}, G., Decouchelle, A., Ballet, J., Hwang, U., Hughes, J. P., Petre, R. 2004a, A\&A, 414, 545

\bibitem[Cassam-Chena\"{i} et al. 2004b]{gcc2}
Cassam-Chena\"{i}, G., Decouchelle, A., Ballet, J., Dubner, G. 2004b, Young Neutron Stars and Their Environments, IAU Symposium, Vol. 218, F. Camilo and B. M. Gaensler, eds.

\bibitem[Castor et al. 1975]{castor}
Castor, J., McCray, R., \& Weaver, R. 1975, ApJ, 200, L107

\bibitem[Chevalier 1982]{chevalier1}
Chevalier, R. A. 1982, ApJ, 258, 790

\bibitem[Chevalier 1983]{chevalier2}
Chevalier, R. A. 1983, ApJ, 272, 765

\bibitem[Chevalier 1999]{chevalier4}
Chevalier, R. A. 1999, ApJ, 511, 798

\bibitem[Crawford et al. 2002]{crawford}
Crawford, F., Pivovaroff, M. J., Kaspi, V. M., \& Manchester, R. N. 2002, Neutrons Stars in Supernova Remnants,
ASP Conference Series, Vol. 9999, P. O. Slane and B. M. Gaensler, eds.

\bibitem[Dame et al. 2001]{dame}
Dame, T. M., Hatmann, D., Thaddeus, P., 2001, ApJ, 547, 792

\bibitem[Decourchelle \& Ballet 1994]{decour1}
Decourchelle, A., \& Ballet, J. 1994, A\&A, 287, 206

\bibitem[Ellison et al. 2000]{ellison1}
Ellison, D. C., Berezhko, E. G., \& Baring, M. G. 2000, ApJ, 540, 292

\bibitem[Ellison et al. 2001]{ellison2}
Ellison, D. C., Slane, P., \& Gaensler, B. M. 2001, ApJ, 563, 191

\bibitem[Enomoto et al. 2002]{enomoto}
Enomoto, R., Tanimori, T., Naito, T. et al. 2002, Nature, 416, 823

\bibitem[Fich et al. 1989]{fich}
Fich,M., Blitz, L., Stark, A.A., 1989, ApJ, 342, 272

\bibitem[Fukui et al. 2003]{fukui}
Fukui, Y., Moriguchi, Y., Tamura, K., et al. 2003, PASJ, 55, L61

\bibitem[Georgelin & Georgelin 1976]{georgelin}
Georgelin, Y. M., Georgelin, Y. P., 1976, A\&A, 49, 57

\bibitem[Hartman et al. 1999]{hartman}
Hartman, R. C., et al. 1999, ApJS, 123, 79

\bibitem[Hester \& Cox 1986]{hester}
Hester, J. J. \& Cox, D. P. 1986, ApJ, 300, 675

\bibitem[Kargaltsev et al. 2002]{kargaltsev}
Kargaltsev, O., Pavlov, G. G., Sanwal, D., \& Garmire, G. P. 2002, ApJ, 580, 1060

\bibitem[Koyama et al. 1995]{koyama1}
Koyama, K., Petre, R., Gotthelf, E. V., Hwang, U., Matsuura, M., Ozaki, M. \& Holt, S. S. 1995, Nature, 378, 255

\bibitem[Koyama et al. 1997]{koyama2}
Koyama, K., Kinugasa, K., Matsuzaki, K. et al. 1997, PASJ, 49, L7

\bibitem[Kennicutt 1984]{kennicutt}
Kennicutt, R. C. 1984, ApJ, 277, 361

\bibitem[Koo et al. 2004]{koo}
Koo, B.-C., Kang, J., \& McClure-Griffiths, N. 2004, Young Neutron Stars and Their Environments, IAU Symposium, Vol. 218, F. Camilo and B. M. Gaensler, eds.

\bibitem[Lazendic et al. 2003]{lazendic1}
Lazendic, J. S., Slane, P. O., Gaensler, B. M., et al. 2003, ApJ, 593, L27

\bibitem[Lazendic et al. 2004]{lazendic2}
Lazendic, J. S., Slane, P. O., Gaensler, B. M., et al. 2004, ApJ, 602, 271

\bibitem[Levenson et al. 1996]{levenson}
Levenson, N. A., Graham, J. R., Hester, J. J., \& Petre, R. 1996, ApJ, 468, 323

\bibitem[McKee et al. 1974]{McKee1}
McKee, C. F., Van Buren, D., \& Lazareff, B. 1984, ApJ, 278, L115

\bibitem[McKee \& Cowie 1975]{McKee2}
McKee, C. F. \& Cowie, L. L. 1975, ApJ, 195, 715

\bibitem[Muraishi et al. 2000]{muraishi}
Muraishi, H., Tanimori, T., Yanagita et al. 2000, A\&A, 354, L57

\bibitem[Pannuti et al. 2003]{pannuti}
Pannuti, T. G., Allen, G. E., Houck, J. C., \& Sturner, S. J. 2003, ApJ, 593, 377

\bibitem[Pavlov et al. 2002]{pavlov}
Pavlov, G. G., Sanwal, D., Garmire, G. P., \& Zavlin, V. E. 2002, ASP Conf. Ser. 271: Neutron Stars in Supernova Remnants, eds. P. O. Slane \& B. M. Gaensler, 247

\bibitem[Pavlov et al. 2002]{pavlov}
Pavlov, G. G., Sanwal, D., \& Teter, M. A. 2004, Young Neutron Stars and Their Environments, IAU Symposium, Vol. 218, F. Camilo and B. M. Gaensler, eds.

\bibitem[Pfeffermann \& Aschenbach 1996]{pfeffermann}
Pfeffermann, E. \& Aschenbach, B. 1996, in Roentgenstrahlung from the Universe, ed. H. H. Zimmermann, J. Tr\"{u}mper, \& H. Yorke (MPE Rep. 263; Garching: MPE), 267

\bibitem[Read \& Ponman 2003]{read}
Read, A. M. \& Ponman, T. J. 2003, A\&A, 409, 395

\bibitem[Reimer \& Pohl 2002]{reimer}
Reimer, O. \& Pohl, M. 2002, A\&A, 390, L43

\bibitem[Reynolds 1998]{reynolds}
Reynolds, S. P., 1998, ApJ, 493, 375

\bibitem[Slane et al. 1999]{slane}
Slane, P., Gaensler, B. M., Dame, T. M. et al. 1999, ApJ, 525, 357, SL99

\bibitem[Slane et al. 2001]{slane2}
Slane, P., Hughes, J. P., Edgar, R. J. et al. 2001, ApJ, 548, 814

\bibitem[Tatematsu et al. 1990]{tatematsu}
Tatematsu, K., Fukui, Y., Iwata, T., Seward, F. D., Nakano, M., 1990, ApJ, 351, 157

\bibitem[Tsurata 1998]{tsurata}
Tsurata, S. 1998, Phys. Rep., 292, 1

\bibitem[Uchiyama et al. 2003]{uchiyama2}
Uchiyama, Y., Aharonian, F. A., \& Takahashi, T. 2003, A\&A, 400, 567

\bibitem[Wang et al. 1997]{wang}
Wang, Z.-R., Qu, Q.-Y., \& Chen, Y. 1997, A\&A, 318, L59

\bibitem[Wardle 1999]{wardle}
Wardle, M. 1999, ApJ, 525, L101

\bibitem[Weaver et al. 1997]{weaver}
Weaver, R., McCray, R., \& Castor, J. 1977, ApJ, 218, 377

\bibitem[Willingale et al. 2001]{willingale}
Willingale, R., Aschenbach, B., Griffits, R. G., Sembay, S., Warwick, R. S., Becker, W., Abbey, A. F., \& Bonnet-Bidaud, J.-M. 2001, A\&A, 365, L212

\bibitem[Wilner et al. 1998]{wilner}
Wilner, D. J., Reynolds, S. P., Moffett, D. A., 1998, AJ, 115, 247

\end{thebibliography}
\end{document}